\documentclass[prd,aps,onecolumn,preprintnumbers, showpacs, nofootinbib,superscriptaddress,notitlepage]{revtex4-1}

\usepackage{bm}
\usepackage{times}
\usepackage{braket}
\usepackage{slashed}
\usepackage{hyperref}
\usepackage{amssymb}
\usepackage{gensymb}

\usepackage{mathrsfs}
\usepackage{amsfonts}
\usepackage{amsmath}
\usepackage{slashed}
\usepackage{array}
\usepackage{verbatim}
\usepackage{epsfig}
\usepackage{graphicx}
\usepackage{color}
\usepackage[dvipsnames]{xcolor}
\usepackage{multirow}
\usepackage{booktabs}
\usepackage{mathrsfs}
\usepackage[normalem]{ulem}

\newcommand{\beq}{\begin{eqnarray}}
\newcommand{\eeq}{\end{eqnarray}}

\newcommand{\non}{\nonumber\\ }

\newcommand{\xsl}{ x \hspace{-2.0truemm}/ }

\newcommand{\zsl}{ z \hspace{-2.0truemm}/ }

\newcommand{\psl}{ p \hspace{-1.8truemm}/ }

\begin{document}

\title{$D_s \to f_0$ form factors and the $D_s^+ \to \left[ \pi\pi \right]_{\rm S} e^+ \nu_e$ decay from light-cone sum rules}

\author{Shan Cheng} \email{Corresponding author: scheng@hnu.edu.cn} 
\author{Shu-Lei Zhang} \email{Corresponding author: zhangshulei@hnu.edu.cn}

\affiliation{School of Physics and Electronics, Hunan University, 410082 Changsha, China.} 
\affiliation{School for Theoretical Physics, Hunan University, 410082, Changsha, China.} 
\affiliation{Hunan Provincial Key Laboratory of High-Energy Scale Physics and Applications, 410082 Changsha, China.}

\date{\today}

\begin{abstract}
In this paper we revisit $D_s \to f_0$ form factors from the light-cone sum rules with the light meson light-cone distribution amplitudes. 
The main motivation of this study is the differential decay width of $D_s \to \left[\pi\pi \right]_{\rm S} e \nu_e$ measured recently by BESIII collaboration 
and the $D_s \to f_0$ form factor extracted under the intermediate resonant model. 
Our result of the differential width of $D_s^+ \to f_0 (\to \left[ \pi\pi \right]_{\rm S}) e^+ \nu_e$ decay 
obtained under the narrow width approximation is a litter bit lower than the data, 
the result obtained under the resonant Flatt\'e model is in consistent with the data while shows a litter bit larger, 
indicating a sizable mixing $\sim 20\degree$ between ${\bar s}s$ and ${\bar u}u+{\bar d}d$ of $f_0$. 
In order to obtain a model independent prediction, we suggest to calculate $D_s \to \left[ \pi\pi \right]_{\rm S}$ form factors 
with the isoscalar scalar dipion light-cone distribution amplitudes. 
Our calculation of $D_s \to \left[ \pi\pi \right]_{\rm S}$ form factors is carried out at the leading twist level due to the finite knowledge of dipion system, 
the result of differential width shows a moderate evolution in contrast to that obtained from the narrow width approximation and the Flatt\'e model, 
revealing a bright prospect to study the four-body leptonic decays of heavy mesons with the dimeson light-cone distribution amplitudes. 

\end{abstract}


\maketitle

\section{introduction}

Weak decays of hadrons containing at least a valence bottom or charm quarks play an important role in the precise examination of standard model (SM) 
and offer one of the best chance for the discovery of new physics (NP) beyond standard model,
in which the semileptonic $D_s$ weak decays provide the clear enviroment to study the structure of light hadrons \cite{El-Bennich:2008rkp}. 
For examples, the semileptonic decay with $D_s \to \eta^{(\prime)}$ transition provides an opportunity to study the $\eta$-$\eta^\prime$ mixing \cite{Bediaga:2003hr,Offen:2013nma}, 
the decay deduced by $D_s \to f_0$ and $D_s \to a_0$ transitions could help us to understand the composition figure of scalar mesons \cite{Cheng:2005nb,DeFazio:2001uc,Ke:2009ed} and the isospin-violated $f_0$-$a_0$ mixing \cite{Aliev:2007uu,BESIII:2021tfk}. 
In this work, we focus on the $D_s^+ \to f_0 e^+ \nu_e$ decay by considering the width effect in the differential width measurements. 

From the spectral analysis, the underlying assignments of scalar mesons like $f_0(980)$ is still not clear. 
Pictures like the tetra-quark \cite{Jaffe:1976ig,Agaev:2017cfz}, the gluonball \cite{Weinstein:1982gc}, the hybrid state \cite{Weinstein:1983gd} 
and the molecule state \cite{Weinstein:1990gu} are all discussed, in which the tetraquark assignment is more favorite nowadays. 
The case is different in the $B$ meson decay where $f_0$ is energetic and the process happens with large recoiling, 
in this case the conventional $q{\bar q}$ assignment is the favorite one because 
the possibility to form a tetra-quark state is power suppressed with comparing to the state of quark pair. 
In the $D_s \to f_0$ decay, one may doubt the $q{\bar q}$ configuration because $f_0$ is not fast moving enough 
so that it has time to pick other $q{\bar q}$ to form a tetraquark. 
This is also our consideration in this paper, 
we take the $q{\bar q}$ configuration in the form factors calculation to check the reliability of energetic picture in charm meson decays. 
From the experiment side, one decade ago, 
CLEO collaboration published the first absolute branching fraction measurement of $D_s$ semileptinic decay including a scalar meson in the final state. 
The result published firstly is ${\cal B}(D_s^+ \to f_0 e^+ \nu_e) \times {\cal B}(f_0 \to \pi^+\pi^-) = \left( 1.3 \pm 0.4 \pm 0.1 \right) \times 10^{-3}$ \cite{CLEO:2009dyb} and subsequently updated to $\left(2.0 \pm 0.3 \pm 0.1 \right) \times 10^{-3}$ in Ref. \cite{CLEO:2009ugx} and also in Ref. \cite{Hietala:2015jqa}.
Recently, the BESIII collaboration has verified the CLEO measurement with much better accuracy, 
the result of product branching fraction is ${\cal B}(D_s^+ \to f_0 e^+ \nu_e) \times {\cal B}(f_0 \to \pi^0\pi^0) = \left(0.79 \pm 0.14 \pm 0.03 \right) \times 10^{-3}$ for the neutral channel \cite{BESIII:2021drk} and ${\cal B}(D_s^+ \to f_0 e^+ \nu_e) \times {\cal B}(f_0 \to \pi^+\pi^-) = \left( 1.72 \pm 0.13 \pm 0.10 \right) \times 10^{-3}$ for the charged channel \cite{BESIII:2023wgr}. 
More important is that BESIII extracted the $D_s \to f_0$ form factor under the Flatt\'e resonant model 
with the data corresponding to an integrated luminosity of $7.33$ fb$^{-1}$, 
the result at the full recoiled point is $f_+(q^2=0) = 0.518 \pm 0.018 \pm 0.036$ \cite{BESIII:2023wgr} with the statistical and systematic errors. 

In this paper, we revisit the $D_s \to f_0(980)$ form factor under the scenario of ${\bar s}s$ and possible mixing with ${\bar u}u+{\bar d}d$. 
With considering the mixing angle $\theta = 20\degree \pm 10\degree$, 
the updated LCSRs calculation of $D_s \to f_0$ form factor 
is in consistent with the one extracted from the differential width of $D_s^+ \to f_0(\left[ \pi\pi \right]_{\rm S}) e^+ \nu_e$ decay under the Flatt\'e model by BESIII. 
while showing a litter bit larger. 
Our calculation is carried out at leading order of strong coupling constant. In order to estimate the next-to-leading-order (NLO) correction, 
we vary the charm quark mass by ${\bar m}_c(m_c) = 1.3 \pm 0.3$ GeV which deduces $20\%-30\%$ additional uncertainty. 
We adopt the Flatt\'e formula to discuss the width effect in semileptonic decay $D_s^+ \to f_0 ( \to \pi^+\pi^- ) e^+ \nu_e$ 
and compare the differential width to the recent measurements at BESIII \cite{BESIII:2023wgr}. 
In order to obtain a model independent prediction, we suggest to calculate $D_s \to \left[ \pi\pi \right]_{\rm S}$ form factors with 
dipion distribution amplitudes ($2\pi$DAs) and compare directly to the measurement without involving resonant. 
The $q^2$ dependence of $D_s^+ \to \left[ \pi^+\pi^- \right]_{\rm S} e^+ \nu_e$ decay width 
indeed shows a different behavior comparing to the result obtained by resonant model.
Our calculation is carried out at leading twist level due to the finite knowledge of $2\pi$DAs, 
more precise measurement is highly anticipated to help us determine the subleading twist $2\pi$DAs. 

The rest of this paper is organized as follows. 
In the next section, the decay constant and LCDAs of scalar isoscalar meson is revisited in the framework of two-point sum rules, 
based on which the $D_s \to f_0$ and $B_s \to f_0$ transition form factors are calculated from the light-cone sum rules in section \ref{sec:Ds2f0-ff}.  
In Section \ref{sec:Ds2pipi-ff}, the chiral even generalized $\pi\pi$ distribution amplitudes is introduced and 
the $D_s \to \left[ \pi\pi \right]_{\rm S}$ form factors are obtained at leading twist level. 
The phenomena related to the recent BESIII measurement is presented in section \ref{sec:Ds2f0-pipi}, 
where the differential width is discussed under the narrow width approximation, Flatt\'e resonant model and the direct $D_s \to \left[ \pi\pi \right]$ transition.
The summary is given in section \ref{sec:summary}.

\section{Light-cone distribution amplitudes of $f_0$}\label{sec:LCDAs}

LCDAs is defined by a nonlocal matrix element sandwiched between vacuum and the onshell hadron state. 
Up to twist three level, the light-cone expansion of scalar meson is 
\beq
&&\langle S(p_1) \vert {\bar q}_{2\beta}(z) q_{1\alpha}(-z) \vert 0 \rangle = \frac{1}{4} \int_0^1 du \, e^{i(2 u - 1) p_1 \cdot z} 
\left\{ \psl_1 \phi(u) + m_S \left[ \phi^s(u) - \frac{\sigma_{\rho\delta} p_1^\rho z^\delta}{6} \phi^\sigma(u) \right] \right\}_{\beta\alpha}, \non
&&\langle S(p_1) \vert {\bar q}_{2\beta}(z) g_s G_{\mu\nu}(vz) \sigma_{\rho\delta} q_{1\alpha}(-z) \vert 0 \rangle = 
\left[ p_{1\mu} \left( p_{1\rho} g_{\nu \delta}^\perp - p_{1\delta} g_{\nu \rho}^\perp \right) \right] 
\int {\cal D} \alpha_i \, \phi_{3S}(\alpha_i) \, e^{ipz(-\alpha_1+\alpha_2+v \alpha_3)} \non 
&& \hspace{5.8cm} - \left[ p_{1\nu} \left( p_{1\rho} g_{\mu \delta}^\perp - p_{1\delta} g_{\mu \rho}^\perp \right) \right] 
\int {\cal D} \alpha_i \, \phi_{3S}(\alpha_i) \, e^{ipz(-\alpha_1+\alpha_2+v \alpha_3)}. 
\label{eq:f0-LCDAs}
\eeq
In the definitions $\phi$ and $\phi^{s,\sigma}$ are the twist two and twist three LCDAs of the ${\bar q}q$ composition, respectively, 
$\phi_{3S}$ is the twist three LCDA in the ${\bar q}gq$ composition. 
Here $v = (u - \alpha_1)/\alpha_3$ and $\alpha_3 = 1- \alpha_1 - \alpha_2$, the measure of three particle integral reads as 
\beq
\int_0^u \, {\cal D} \alpha_i = \int_0^u d \alpha_1 \int_0^{1-u} \frac{d \alpha_2 }{1-\alpha_1-\alpha_2}.
\eeq

The leading twist and two particle twist three LCDAs are normalized to the vector and scale-dependent scalar decay constants, 
which are defined by the local matrix elements deduced by the vector and scalar currents, respectively, 
\beq
&&\int_0^1 du \, \phi(u) = f_S, \qquad \langle S(p_1) \vert {\bar q}_{2} \gamma_\mu q_1 \vert 0 \rangle = f_S p_{1 \mu}, 
\label{eq:LCDAs-norm-1}\\
&&\int_0^1 du \, \phi^{(s/\sigma)}(u) = {\bar f}_S, \qquad \langle S(p_1) \vert {\bar q}_2 q_1 \vert 0 \rangle = {\bar f}_S m_S. 
\label{eq:LCDAs-norm-2}
\eeq
For the neutral scalar meson like $f_0, a_0$ which could not be produced via the vector current, 
$f_{S} = 0$ due to the charge conjugate invariance or the conservation of vector current. 
This is also implied in the relation 
\beq
{\bar f}_S = \mu_S f_s, \qquad \mu_S \equiv \frac{m_S}{m_{q_2}(\mu) - m_{q_1}(\mu)},
\eeq 
from which we can see that $f_S$ vanishes in the ${\rm SU(3)}$ or isospin limit.

\subsection{The scalar decay constant}\label{sssec:decaycons}

To calculate the scalar coupling, we consider the following correlation function
\beq
\Pi(q) = i \int d^4x e^{i q \cdot x} \langle 0 \vert {\rm T} \{ {\bar q}_2(x)q_1(x), {\bar q}_1(0)q_2(0) \} \vert 0 \rangle.
\label{eq:correlator-coupling}
\eeq
The QCD sum rule (QCDSR) of the neutral scalar coupling 
is quoted as \cite{Govaerts:1986ua} 
\beq
&&m_S^2 {\bar f}_S^2(\mu) e^{- m_S^2/M^2} = I_0^{\rm pert}(s_0,M^2) +  \langle {\bar q}q \rangle I_0^{\langle {\bar q}q \rangle}(M^2) 
+ \langle \alpha_s G^2 \rangle I_0^{\langle \alpha_s G^2  \rangle}(M^2, \mu) \non
&& \hspace{2.8cm}+ \langle g_s {\bar q} \sigma T G q \rangle I_0^{\langle g_s {\bar q} \sigma T G q \rangle}(M^2) 
+ \langle g_s{\bar q}q \rangle^2 I_0^{\langle g_s{\bar q}q \rangle^2}(M^2) 
+ \langle g_s^2{\bar q}q \rangle^2 I_0^{\langle g_s^2{\bar q}q \rangle^2}(M^2,\mu) 
\label{eq:sr-decaycons}
\eeq
with the weighted functions 
\beq
&&I_0^{\rm pert}(s_0,M^2) = \frac{3M^4}{8\pi^2} 
\left[ 1 + \frac{\alpha_s(\mu)}{\pi} \left( \frac{17}{3} + 2 \frac{I(1)}{f(1)} - 2 \ln \frac{M^2}{\mu^2} \right) f(1) \right], \non  
&&I_0^{\langle {\bar q}q \rangle}(M^2) = 3m_q, \qquad
I_0^{\langle \alpha_s G^2 \rangle}(M^2, \mu) = \frac{1}{8\pi}, \non
&&I_0^{\langle g_s {\bar q} \sigma T G q \rangle}(M^2) = - \frac{m_q}{M^2}, \qquad
I_0^{\langle g_s{\bar q}q \rangle^2}(M^2) = \frac{2 \pi \alpha_s}{3M^2}, \qquad
I_0^{\langle g_s^2{\bar q}q \rangle^2}(M^2,\mu) = \frac{\pi \alpha_s}{M^2},
\label{eq:sr-decaycons-tra}
\eeq
and the functions $I(1) = \int_{e^{-s_0/M^2}}^1 (\ln t) \ln(- \ln t) dt$ and $f(1) = 1- e^{-s_0/M^2} \left( 1 + s_0/M^2 \right)$. 
The renormalization group equations of scalar coupling and vacuum condensations are \cite{Yang:1993bp,Lu:2006fr}
\beq
&&{\tilde f}_S(\mu) = \tilde{f}_S(\mu_0) L(\mu,\mu_0), \qquad m_q(\mu) = m_q(\mu_0) L^{-1}(\mu,\mu_0), \qquad 
{\langle {\bar q}q(\mu) \rangle} = {\langle {\bar q}q(\mu_0) \rangle} L(\mu,\mu_0), \non
&&\langle g_s {\bar q} \sigma T G q(\mu) \rangle = \langle g_s {\bar q} \sigma T G q(\mu_0) \rangle L^{-1/6}(\mu,\mu_0), \qquad
\langle g_s{\bar q}q(\mu) \rangle^2 = \langle g_s{\bar q}q(\mu_0) \rangle^2 L(\mu,\mu_0). 
\eeq
Here $L(\mu,\mu_0) = \left[ \alpha_s(\mu_0)/\alpha_s(\mu) \right]^{4/b}$. 
In the numerics, we take $\alpha_s (1 \, {\rm GeV}) = 0.47$ corresponding to the world average $\alpha_s(m_Z) = 0.118$ 
and the follow vacuum condensates \cite{Ioffe:2002ee} where the acquiescent scale is $1$ GeV. 
\beq
&&\langle {\bar q}q \rangle = - 0.0156 \, {\rm GeV}^3, \qquad
\langle {\bar s}s \rangle = 0.8 \langle {\bar q}q \rangle, \qquad
\langle \alpha_s G^2 \rangle = 0.012 \pi \, {\rm GeV}^4, \qquad 
m_0^2 = 0.8, \non
&&\langle g_s {\bar q} \sigma T G q \rangle = m_0^2 \langle {\bar q}q \rangle , \qquad
\langle g_s{\bar q}q \rangle^2 = \frac{-16}{9} \langle {\bar q}q \rangle^2 , \qquad 
\langle g_s^2{\bar q}q \rangle^2 = \frac{-16}{3} \langle {\bar q}q \rangle^2 .
\eeq

The scalar coupling is also studied by the sum rules within the back ground field (BFTSR) \cite{Wu:2022qqx}, 
where the perturbative term, the quark condensate, dimension-$4$ gluon condensate and dimension-$6$ quark-gluon condensate 
are the same as in the traditional one as shown in Eq. (\ref{eq:sr-decaycons-tra}), 
while the dimension-$6$ four quark condensates are different.  
\beq
I_0^{\langle g_s{\bar q}q \rangle^2}(M^2) = - \frac{8}{27M^2}, \qquad
I_0^{\langle g_s^2{\bar q}q \rangle^2}(M^2,\mu) = \frac{2+\kappa^2}{486 \pi^2 M^2} \left[ 35 + 6 \ln \frac{M^2}{\mu^2} \right]  .
\label{eq:sr-decaycons-bgf}
\eeq
Meanwhile, the values of the non-perturbative vacuum condensates appearing in the BFTSR are also different from them in the traditional ones.  
\beq
&&\langle {\bar q}q \rangle = -0.0242 \, {\rm GeV}^3, \qquad 
\langle {\bar s}s \rangle = 0.8 \langle {\bar q}q \rangle, \qquad
\langle \alpha_s G^2 \rangle = 0.038 \, {\rm GeV}^4, \qquad \non
&&\langle g_s {\bar q} \sigma T G q \rangle = - 0.0193 \, {\rm GeV}^5, \qquad 
\langle g_s{\bar q}q \rangle^2 = 2.082 \times 10^{-3} \, {\rm GeV}^6, \qquad \non
&&\langle g_s^2{\bar q}q \rangle^2 = 7.420 \times 10^{-3 } \, {\rm GeV}^6, \qquad
\langle g_s^2 f G^2 \rangle = 0.045 \, {\rm GeV}^6.
\eeq

\begin{table}[tb]
\begin{center}
\begin{tabular}{l |  c  c | c c  c }
\hline\hline
{\rm Sum rules} \quad & \quad {\rm QCDSR}  \quad & \quad {\rm BFTSR}  \quad & \quad {\rm QCDSR} \cite{Cheng:2005nb} \quad & 
\quad {\rm QCDSR} \cite{Lu:2006fr} \quad & \quad {\rm BFTSR} \cite{Han:2013zg}\quad \non
\hline
$m_{f_0(980)}$({\rm MeV}) \quad & \multicolumn{2}{|c|}{ \quad $990 \pm 50$ ({\rm input}) \quad } & 
\quad $990 \pm 50$ \quad & \quad --- \quad & \quad --- \quad \non
$\tilde{f}_{f_0(980)}$({\rm MeV}) \quad & \quad $335^{+9}_{-12}$ \quad & \quad $331^{+9}_{-12}$  \quad & 
\quad $370 \pm 20$ \quad & \quad --- \quad  & \quad --- \quad \non
$m_{f_0(1500)}$({\rm MeV}) \quad & \quad --- \quad & \quad --- \quad & \quad $1500 \pm 100$ \quad & 
\quad $[1640, 1730]$ \quad   &  \quad $[1563, 1706]$ \quad \non
$\tilde{f}_{f_0(1500)}$({\rm MeV}) \quad &\quad --- \quad & \quad --- \quad & \quad $- (255 \pm 30)$ \quad & 
\quad $[369,391]$ \quad &   \quad $[374,378]$ \quad \non
\hline
$M^2$({\rm GeV}$^2$) \quad & \multicolumn{2}{|c|}{ \quad $1.6 \pm 0.1$ \quad } & \quad $[1.1, 1.6]$ \quad & \quad $1.1 \pm 0.1$ \quad & \quad $2.0 \pm 0.2$ \quad \non
$s_0$({\rm GeV}$^2$) \quad &  \multicolumn{2}{|c|}{ \quad $2.2 \pm 0.2$  \quad } & \quad $5.0 \pm 0.3$ \quad &  \quad $6.5 \pm 0.3$ \quad & \quad $6.5 \pm 0.3$ \quad \non
\hline\hline
\end{tabular}
\caption{The scalar decay constant of $f_0$ obtained from QCD sum rules.}
\label{tab-f0-mass-decaycons}
\end{center}
\end{table}

The Borel mass $M^2$ is usually chosen by a priori criterion that the contribution from high dimension condensates is no more than twenty percents, 
and simultanously the contribution from high excited states and continuum spectral is smaller than thirty percents, 
so it is a compromise between the unitary interpolation from the hadron summing and the operator production expansion from the QCD calculation. 
The threshold value $s_0$ is close to the outset of the first excited state with the same quantum number and 
determined by the maximal stability of physical quantities once the Borel mass has been set down. 
Under the statement of ${\bar f}^s_{\sigma} = 0$ \cite{Cheng:2019tgh}, we consider the $f_0$ as the ground state with ${\bar s}s$ component 
in the sum rules Eq. (\ref{eq:correlator-coupling}) with $q_1=q_2=s$. 
Taking $M^4 \partial/\partial M^2 \ln$ to both sides of Eq. (\ref{eq:sr-decaycons}), we obtain the sum rules of $m_{f_0}$ 
which is then fixed at the PDG value $990 \pm 50 \, {\rm MeV}$. 
With this input, we ultimately find the optimal choice of Borel mass and threshold value are the same in the two types sum rules of $m_{f_0}$ as shown in 
Eq. (\ref{eq:sr-decaycons-tra}) and Eq. (\ref{eq:sr-decaycons-bgf}). 
This agreement is not a coincidence but a rational result since the only difference between the QCDSR and BFTSR 
is the expression of terms proportional to dimension-$6$ four-quark condensate, which is highly power suppressed. 

\begin{figure}[t]
\begin{center}
\resizebox{0.45\textwidth}{!}{
\includegraphics{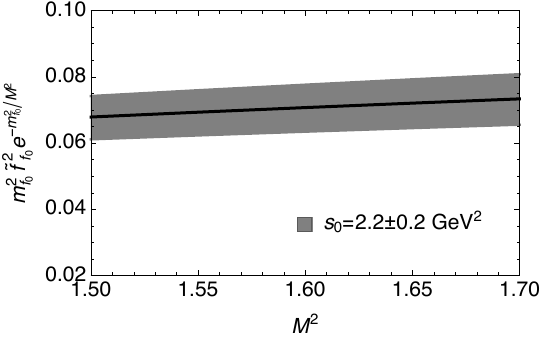}} 
\hspace{8mm}
\resizebox{0.45\textwidth}{!}{
\includegraphics{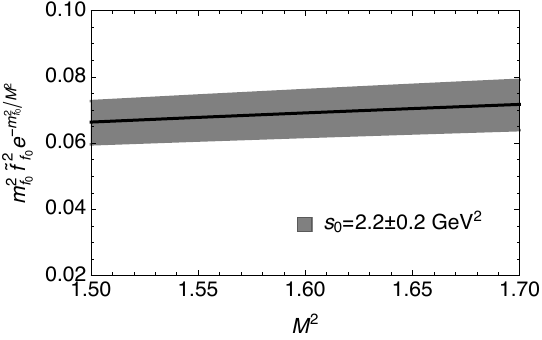}} \non
\end{center}
\vspace{-8mm}
\caption{The Borel mass dependence of $m^2_{f_0} \tilde{f}^2_{f_0} e^{-m^2_{f_0}/M^2}$ obtained from QCDSR (left) and BFTSR (right).}
\label{fig:mff0-M2}       
\end{figure}

In table \ref{tab-f0-mass-decaycons}, we show the sum rules result of scalar decay constant at the default scale $1$ GeV. 
For the sake of comparison, we also present the previous sum rules result \cite{Cheng:2005nb} 
where both the ground state $f_0$ and the first excited state $f_0(1500)$ are taken in to account. 
The QCDSR is is also studied under the statement that $f_0(1500)$ is the ground state with conventional ${\bar s}s$ component 
while $f_0(980)$ is more like a tetraquark state \cite{Lu:2006fr}, and the similar BFTSRs study is carried out in Ref. \cite{Han:2013zg}.
In figure \ref{fig:mff0-M2}, we show the Borel mass dependence of the sum rules of scalar decay constant as shown in Eq. (\ref{eq:sr-decaycons}), 
in which the uncertainty band corresponds to different values of $s_0$.

\subsection{Leading twist light-cone distribution amplitudes}\label{sssec:LCDAs-t2}

The leading twist LCDA is usually expanded in terms of the gegenbauer polynomials 
\beq
\phi(u, \mu) = {\bar f}_S 6 u {\bar u} \sum_{n=0}^\infty a_n^{\rm t2}(\mu) C_n^{3/2}(2u-1) , 
\label{eq:phi-t2}
\eeq
from which the multiplicatively renormalization coefficients can be written by 
\beq
a_n(\mu) = \frac{1}{{\bar f}_S} \frac{2(2n+3)}{3(n+1)(n+2)} \int_0^1 C_n^{3/2}(2u-1) \phi(u, \mu) du.  
\label{eq:phi-t2-an}
\eeq
Substituting Eq. (\ref{eq:phi-t2}) into the normalization condition in Eq. (\ref{eq:LCDAs-norm-1}), 
we obtain $a_{n = {\rm even}} \propto 1/\mu_S$ and hence the even coefficients are zero for the neutral scalar mesons. 
Taking in to account the expansion of gegenbauer polynomials, the renormalization coefficients is rewritten by means of the moments $\langle \zeta_n \rangle$  
\beq
\langle \zeta^{q}_{n} \rangle =  \frac{1}{{\bar f}_S} \int_0^1 (2u-1)^n \phi(u, \mu) du.
\label{eq:phi-t2-an}
\eeq
The first two renormalization coefficients are related to the moments by 
\beq
a_1(\mu) = \frac{5}{3} \langle \zeta^{q}_1(\mu) \rangle, \qquad 
a_3(\mu) = \frac{21}{4} \langle \zeta^{q}_3(\mu) \rangle - \frac{9}{4} \langle \zeta^{q}_1(\mu) \rangle.
\eeq
Here $q=u,d,s$ is the quark component of scalar meson. 
The renormalization group equation of the gegenbauer coefficients read as 
\beq
a_n(\mu) = a_n(\mu_0) \left[ \frac{\alpha_s(\mu_0)}{\alpha_s(\mu)} \right]^{-\left( \gamma_n^{(0)} \right)/b} L^{-1}(\mu,\mu_0)
\eeq
with the one-loop anomalous dimension 
\beq
\gamma_n^{(0)} 
= C_F \left[ 3 + \frac{2}{(n+1)(n+2)} - 4 \psi(n+2) - 4 \gamma_E \right], 
\eeq
in which $b = (11 N_c - 2 n_f)/3, C_F = (N_c^2-1)/(2N_c)$.

We consider the correlation function 
\beq
\Pi_n(z,q) = i \int d^4x e^{iq \cdot x} \langle 0 \vert {\rm T} \{ {\bar q}_{2}(x) \zsl \left( i z \cdot \overleftrightarrow{D} \right)^n q_1(x), {\bar q}_1(0) q_2(0) \} \vert 0 \rangle.
\label{eq:correlator-moment}
\eeq
In the deep Euclidean region $q^2 \ll 0$, it can be evaluated directly by applying the OPE technique
\beq
&&\Pi_n(z,q) = i \int d^4x e^{iq \cdot x} \left\{ - {\rm Tr.} \langle 0 \vert S_0^{q_2}(0,x) \zsl \left( i z \cdot \overleftrightarrow{D} \right)^n S_0^{q_1}(x,0) \} \vert 0 \rangle 
+ \langle 0 \vert {\bar q}_2(x)q_2(0) \zsl \left( i z \cdot \overleftrightarrow{D} \right)^n S_0^{q_1}(x,0) \} \vert 0 \rangle \right. \non
&&\left. \hspace{3.7cm} + {\rm Tr.} \langle 0 \vert S_0^{q_2}(0,x) \zsl \left( i z \cdot \overleftrightarrow{D} \right)^n {\bar q}_1(x)q_1(0) \} \vert 0 \rangle + \cdots \right\}.
\eeq 
When $q^2$ shifts from deeply negative to positive, the ${\bar q}q$ state begins to form hadrons and 
the correlation function can be expressed by the sum of contributions from all possible intermediate states with appropriate subtractions. 
Writing the hadron representation in the dispersion relation and isolating the ground state contribution, for $q^2 > 0$ we have 
\beq
\Pi_n(z,q) = \frac{1}{\pi} \int_0^\infty ds \frac{{\rm Im} \Pi^{\rm had}_n(s,q^2)}{s - q^2} 
= {\bar f}_S^2 m_S \langle \zeta_n \rangle + \frac{1}{4\pi^3} \int_{s_0}^\infty ds \frac{3m_q}{(n+2) (s - q^2)},
\eeq
in which the relationships 
$\langle 0 \vert {\bar q}_{2} \zsl ( i z \cdot \overleftrightarrow{D} )^n q_1 \vert 0 \rangle = (z \cdot q)^{n+1} {\bar f}_S \langle \zeta_n \rangle$ 
and $\langle 0 \vert {\bar q}_{1} q_2 \vert S \rangle = m_S {\bar f}_S \langle \zeta^s_0 \rangle$ are implied. 
We take the convention $\langle \zeta^s_0 \rangle = 1$ in the following.

After implement the quark-hadron duality and the Borel transformation, 
the QCDSRs of leading twist LCDA moments ($n = {\rm odd}$) is 
\beq
\langle \zeta_n(\mu) \rangle = I_n^{\rm pert}(s_0,M^2) + \langle {\bar q}q \rangle I_n^{\langle {\bar q}q \rangle}(M^2) 
+ \langle g_s {\bar q} \sigma T G q \rangle I_n^{\langle g_s {\bar q} \sigma T G q \rangle}(M^2) 
+ \langle \alpha_s G^2 \rangle \langle {\bar q}q \rangle I_n^{\langle \alpha_s G^2  \rangle \langle {\bar q}q \rangle}(M^2, \mu).  
\label{eq:sr-ant2-qcdsr}
\eeq
The weighted functions $I_n$ describe the contributions from perturbative and various condensate effects. 
For the neutral scalar meson $q_1=q_2=q$, the result up to dimension six is quoted \cite{Cheng:2005nb} as 
\beq
&&I_n^{\rm pert}(s_0,M^2) = - \frac{m_q}{m_S{\bar f}_S^2} e^{m_S^2/M^2} \frac{3 M^2}{8\pi^2(n+2)} \left(1-e^{-s_0/M^2}\right), \non
&&I_n^{\langle {\bar q}q \rangle}(M^2) = \frac{2}{m_S{\bar f}_S^2} e^{m_S^2/M^2}, \non
&&I_n^{\langle g_s {\bar q} \sigma T G q \rangle}(M^2) = \frac{1}{m_S{\bar f}_S^2} e^{m_S^2/M^2} \frac{10n-3}{12 M^2},  \non
&&I_n^{\langle \alpha_s^2 G^2  \rangle \langle {\bar q}q \rangle}(M^2, \mu) = \frac{1}{m_S{\bar f}_S^2} e^{m_S^2/M^2} \frac{4 \pi n(4n-5)}{18 \pi M^4} .
\label{eq:ant2-qcdsr}
\eeq

The leading twist LCDA moments are also studied under the BFTSR \cite{Wu:2022qqx} and the result is
\beq
&&\langle \zeta_{n={\rm odd}}(\mu) \rangle = I_n^{\rm pert}(s_0,M^2) + \langle {\bar q}q \rangle I_n^{\langle {\bar q}q \rangle}(M^2) 
+ \langle g_s {\bar q} \sigma T G q \rangle I_n^{\langle g_s {\bar q} \sigma T G q \rangle}(M^2) 
+ \langle \alpha_s G^2 \rangle I_n^{\langle \alpha_s G^2  \rangle}(M^2, \mu) \non
&& \hspace{1cm}+ \langle \alpha_s G^2 \rangle \langle {\bar q}q \rangle I_n^{\langle \alpha_s G^2 \rangle \langle {\bar q}q \rangle}(M^2) 
+ \langle g_s{\bar q}q \rangle^2 I_n^{\langle g_s{\bar q}q \rangle^2}(M^2) 
+ \langle g_s^2{\bar q}q \rangle^2 I_n^{\langle g_s^2{\bar q}q \rangle^2}(M^2,\mu) 
+ \langle g_s^3 f G^3 \rangle I_n^{\langle g_s^3 f G^3 \rangle}(M^2,\mu).
\label{eq:sr-ant2-bftsr}
\eeq
The weighted functions associated to the perturbative and quark condensate terms are the same as in the QCDSR. 
For other nonperturbative condensates with $n \leqslant 1$, they are
\beq
&&I_n^{\langle g_s {\bar q} \sigma T G q \rangle}(M^2) = - \frac{1}{m_S{\bar f}_S^2} e^{m_S^2/M^2} \frac{4n}{3 M^2},  \non
&&I_n^{\langle \alpha_s G^2  \rangle}(M^2, \mu) = \frac{2m_q}{m_S{\bar f}_S^2} \frac{e^{m_S^2/M^2}}{48 \pi M^2} 
\left\{ - 12n \ln \frac{M^2}{\mu^2} - 6(n+2) + \Theta(n-1) \left[ -4n \ln \frac{M^2}{\mu^2} + 3 \tilde{\psi}(n) - \frac{6}{n} \right] \right\}, \non
&&I_n^{\langle g_s{\bar q}q \rangle^2}(M^2) = \frac{m_q}{m_S{\bar f}_S^2} e^{m_S^2/M^2} \frac{4(n+3)}{81 M^4},  \non
&&I_n^{\langle g_s^2{\bar q}q \rangle^2}(M^2, \mu) =  - \frac{2m_q}{m_S{\bar f}_S^2} e^{m_S^2/M^2} \frac{2+\kappa^2}{3888\pi^2 M^4} 
\left\{4(n+5) + \delta^{0n} \left[ 24 \ln \frac{M^2}{\mu^2} -148 \right] + \delta^{1n} \left[ -128 \ln \frac{M^2}{\mu^2} -692 \right] \right. \non
&& \left. \hspace{2.4cm} + \Theta(n-1) \left[ -8(6n^2+34n) \ln \frac{M^2}{\mu^2} + 4n \tilde{\psi}(n) - 2(6n^2+96n+212) \right] \right\}, \non
&&I_n^{\langle g_s^3 f G^3 \rangle}(M^2,\mu) = - \frac{2m_q}{m_S{\bar f}_S^2} e^{m_S^2/M^2} \frac{1}{384\pi^2 M^4} 
\left\{ \delta^{1n} \left[ 24 \ln \frac{M^2}{\mu^2} + 84 \right] \right. \non
&& \left. \hspace{2.4cm} + \Theta(n-1) \left[ 4n(3n-5) \ln \frac{M^2}{\mu^2} + 2(2n^2+5n-13) \right] \right\}.
\label{eq:ant2-bftsr}
\eeq
Here $ \tilde{\psi}(n) = \psi((n+1)/2) - \psi(n/2) + (-1)^n \ln 4$, $\psi(n+2) = \sum_{j=1}^{n+1} 1/j - \gamma_E$, $\kappa = 0.74 \pm 0.03$.

\begin{figure}[t]
\begin{center}
\resizebox{0.45\textwidth}{!}{
\includegraphics{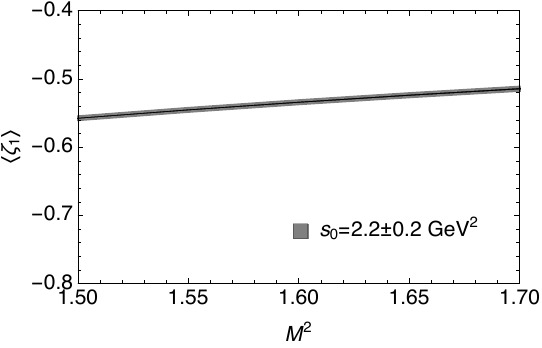}} 
\hspace{8mm}
\resizebox{0.45\textwidth}{!}{
\includegraphics{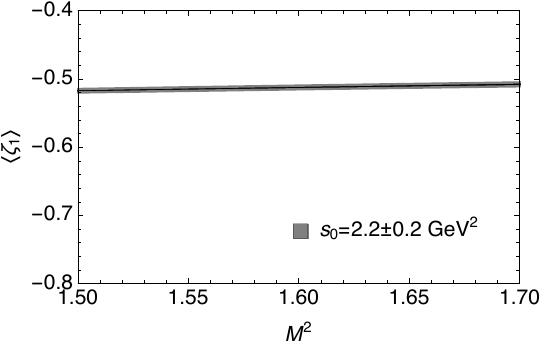}} \non
\end{center}
\vspace{-6mm}
\caption{The Borel mass dependence of the first moment obtained from the QCDSR (left)  and BFTSR (right).}
\label{fig:a1t2-M2}       
\end{figure}

In figure \ref{fig:a1t2-M2} we show the Borel mass dependence of the first moment obtained from sum rules. 
Because the moments are dominated by the nonperturbative contributions, 
we see that $\langle \zeta_1 \rangle $ is not sensitive to the threshold value $s_0$. 
In table \ref{tab-f0-a1} we show the result of the first gegenbauer coefficient, where the first and second errors arise from $M^2$ and $s_0$, respectively. 
For the sake of comparison, we also present the result obtained from previous sum rules\footnote{In the last column, 
we take the ${\rm SU(3)}$ asymptotic and compare directly to the result obtained for the isovector scalar meson $a_0(980)$ \cite{Han:2013zg}.}. 
Our result of the first gegenbauer coefficient $a_1$ obtained from the QCDSR is consist with the previous sum rules determination \cite{Cheng:2005nb} , 
the result is almost the same as that obtained from the BFTSR with the same sum rules parameters. 
 
\begin{table}[t]
\vspace{4mm}
\begin{tabular}{l |  c  c | c c  c }
\hline\hline
{\rm Sum rules} \quad & \quad {\rm QCDSR}  \quad & \quad {\rm BFTSR}  \quad & \quad {\rm QCDSR} \cite{Cheng:2005nb} \quad & 
\quad {\rm QCDSR} \cite{Wang:2008da} \quad & \quad {\rm BFTSR} \cite{Wu:2022qqx}\quad \non
\hline
$a_1(f_0(980))$ & $-0.891^{+0.039+0.004}_{-0.033-0.004} $ \quad & $-0.855^{+0.039+0.004}_{-0.033-0.005}$ & $-0.78 \pm 0.08$ & 
 ---  &  $-0.51 \pm 0.07$ \non
$a_1(f_0(1500))$ & --- & --- & $0.80 \pm 0.40$ & $-0.48 \pm 0.11$  & ---  \non
\hline
$M^2$({\rm GeV}$^2$)  & \multicolumn{2}{c|}{ $1.6 \pm 0.1$ } & \multicolumn{2}{c}{ $[1.1, 1.6]$} & $[2.5, 3.5]$ \non
$s_0$({\rm GeV}$^2$) &  \multicolumn{2}{c|}{ $2.2 \pm 0.2$ } & \multicolumn{2}{c}{ $5.0 \pm 0.3$ } & $7.0$ \non
\hline\hline
\end{tabular}
\caption{The first gegenbauer coefficients obtained from QCD sum rules.}
\label{tab-f0-a1}
\end{table}

\subsection{Subleading twist light-cone distribution amplitudes}\label{sssec:LCDAs-t3}

Twist three light-cone distribution amplitudes associated with two particle composition are expanded also in terms of gegenbauer polynomials \cite{Wang:2008da}
\beq
&&\phi^s(u,\mu) = {\bar f}_S \left[ 1 + \sum_{n=1}^\infty a_n^{s}(\mu) C_n^{1/2}(2u-1) \right], \non
&&\phi^\sigma(u,\mu) = {\bar f}_S 6u {\bar u} \left[ 1 + \sum_{n=1}^\infty a_n^{s}(\mu) C_n^{3/2}(2u-1) \right].
\label{eq:phi-t3}
\eeq 
The scalar and tensor moments defined in
\beq
\langle \zeta_n^{s/\sigma} \rangle = \int_0^1 du (2u-1)^n \phi^{s/\sigma}(u,\mu) 
\label{eq:phi-t3-moments}
\eeq
relate to the gegenbauer coefficients $a_n^{s/\sigma}$ by
\beq
&&a_1^s = 3 \langle \zeta_1^s \rangle, \quad a_2^s = \frac{5}{2} \left[ 3 \langle \zeta_2^s \rangle - 1 \right], \quad 
a_4^s = \frac{9}{8} \left[35 \langle \zeta_4^s \rangle - 30 \langle \zeta_2^s \rangle + 3 \right], \non
&&a_1^\sigma = \frac{5}{3} \langle \zeta_1^\sigma \rangle, \quad a_2^\sigma = \frac{7}{12} \left[ 5 \langle \zeta_2^\sigma \rangle - 1 \right], \quad 
a_4^\sigma = \frac{11}{24} \left[21 \langle \zeta_4^\sigma \rangle - 14 \langle \zeta_2^\sigma \rangle + 1 \right]. 
\label{eq:phi-t3-coeff}
\eeq

We consider two correlation functions 
\beq
&&\Pi_n^s(z,q) = i \int d^4x \, e^{iq \cdot x} \langle 0 \vert {\rm T} 
\left\{ {\bar q}_2(x) \left( i z \cdot \overleftrightarrow{D} \right)^n q_1(x), {\bar q}_1(0) q_2(0) \right\} \vert 0 \rangle
= - \left(z \cdot q \right)^n I_n(q^2), 
\label{eq:correlator-moment-s}\\
&&\Pi_n^\sigma(z,q) = i \int d^4x \, e^{iq \cdot x} \langle 0 \vert {\rm T} 
\left\{ {\bar q}_2(x) \sigma_{\mu\nu} \left( i z \cdot \overleftrightarrow{D} \right)^n q_1(x), {\bar q}_1(0) q_2(0) \right\} \vert 0 \rangle 
 = i \left( q_\mu z_\nu - q_\nu z_\mu \right) \left( z \cdot q \right)^n I_n^\sigma(q^2). 
\label{eq:correlator-moment-sigma}
\eeq
The dispersion relation representation of the correlation functions in the physical regions reads as 
\beq
\Pi_n^{s/\sigma}(z,q) = \frac{1}{\pi} \int ds \frac{{\rm Im} \Pi_n^{s/\sigma, {\rm had}}(s,q^2)}{s-q^2}. 
\eeq
With the definitions of local matrix elements in terms of moments
\beq
&&\langle 0 \vert {\bar q}_2(x) \left( i z \cdot \overleftrightarrow{D} \right)^n q_1(x) \vert S(q) \rangle 
= m_S {\bar f}_S \left( q \cdot z \right)^n \langle \zeta_n^s \rangle, \non
&&\langle 0 \vert {\bar q}_2(x) \sigma_{\mu\nu} \left( i z \cdot \overleftrightarrow{D} \right)^{n+1} q_1(x) \vert S(q) \rangle 
= - i \frac{n+1}{3} m_S {\bar f}_S \left( q_{\mu}z_\nu - q_{\nu}z_\mu \right) \left( q \cdot z \right)^n \langle \zeta_n^\sigma \rangle,  
\eeq
we obtain the imaginary parts for the neutral scalar mesons\footnote{We here concentrate on the correlation functions with even $n$ 
since the odd ones result to zero moments for the neutral meson due to the $C$-parity conservation.}, 
\beq
&&\frac{1}{\pi} {\rm Im} I^{s, {\rm had}}_{n = {\rm even}} = - \delta(q^2 - m_S^2) m_S^2 {\bar f}_S^2 \langle \zeta_{n= {\rm even}}^s \rangle 
+ \Theta(q^2 - s_0^s) \frac{3 \left( -q^2 + 4m_q^2 \right)}{8 \pi^2} \frac{1}{n+1}, \non
&&\frac{1}{\pi} {\rm Im} I^{\sigma, {\rm had}}_{n = {\rm even}} = - \delta(q^2 - m_S^2) m_S^2 {\bar f}_S^2 \langle \zeta_{n= {\rm even}}^\sigma \rangle  \frac{n+1}{3}
+ \Theta(q^2 - s_0^s) \frac{3 \left( -q^2 + 2 m_q^2 
\right)}{8 \pi^2} \frac{1}{n+3}.
\label{eq:correlator-moment-t3-Im}
\eeq

In the deep Euclidean region, the correlation functions can be evaluated using operator product expansion at quark level. 
After applying the quark-hadron duality to math the result of $I_n^{s/\sigma}(q^2)$ obtained from the hadron interpolating and OPE calculation,  
for the neutral scalar mesons, the Borelization result of the second moments are \cite{Lu:2006fr} 
\beq
&&-m_S^2 {\bar f}_S^2 e^{-m_S^2/M^2} \langle \zeta_{n = {\rm even}}^s \rangle 
= - \frac{3}{8 \pi^2} \frac{1}{n+1} \int_0^{s_0^s} s e^{-s/M^2} ds - \langle \alpha_s G^2 \rangle \frac{1}{8\pi} 
- \langle {\bar q} q \rangle (n+3) m_q  \non
&&\hspace{2cm}+ \langle g_s {\bar q} \sigma T G q \rangle \frac{(4 n^2 + 23n + 12) m_{q}}{12} \frac{1}{M^2} 
+ 4\pi \alpha_s \langle {\bar q} q \rangle^2 \frac{(-8n^2 + 16n + 150)}{81} \frac{1}{M^2},
\label{eq:zeta-s-even-qcdsr} \\
&&-\frac{1}{3} m_S^2 {\bar f}_S^2 e^{-m_S^2/M^2} \langle \zeta_{n = {\rm even}}^\sigma \rangle 
= - \frac{3}{8 \pi^2} \frac{1}{(n+1)(n+3)} \int_0^{s_0^\sigma} s e^{-s/M^2} ds 
- \langle \alpha_s G^2 \rangle \frac{1}{24 \pi} \frac{1}{n+1} - \langle {\bar q} q \rangle m_q  \non
&&\hspace{2cm} + \frac{1}{n+1} \frac{1}{M^2} \left\{ \langle g_s {\bar q} \sigma T G q \rangle \frac{(4n^2+7n+5) m_q}{12} 
- 4\pi \alpha_s \langle {\bar q} q \rangle^2 \left[ \frac{8n^2+9n-35)}{162} + \frac{2 \delta_{n0}}{9} \right]  \right\},
\label{eq:zeta-sigma-even-qcdsr}
\eeq

Besides the QCDSR result, the BFTSR also applied to calculate the scalar and tensor moments 
with the accuracy up to the linear terms of quark mass contributions \cite{Han:2013zg}. 
\beq
&&-m_S^2 {\bar f}_S^2 e^{-m_S^2/M^2} \langle \zeta_{n = {\rm even}}^s \rangle 
= - \frac{3}{8 \pi^2} \frac{1}{n+1} M^4 
+ \frac{3}{8 \pi^2} \frac{e^{-s_0^s/M^2}}{n+1} \left[ M^4 + s_0^s M^2 - 2 (n+2) m_q^2  M^2 \right]  \non
&&\hspace{2cm} - \langle \alpha_s G^2 \rangle \frac{1}{8\pi} 
- \langle {\bar q} q \rangle (n+3) m_q + \langle g_s^3 f G^3 \rangle \frac{n}{96 \pi^2}  \frac{1}{M^2}  \non
&&\hspace{2cm}+ \langle g_s {\bar q} \sigma T G q \rangle \frac{(8 n^2 +13n - 18) m_{q}}{18} \frac{1}{M^2} 
+ 4\pi \alpha_s \langle {\bar q} q \rangle^2 \frac{(-4n^2 - 14n +168 )}{81} \frac{1}{M^2}, 
\label{eq:zeta-s-even-bftsr} \\
&&-\frac{1}{3} m_S^2 {\bar f}_S^2 e^{-m_S^2/M^2} \langle \zeta_{n = {\rm even}}^\sigma \rangle 
= \frac{3}{8 \pi^2} \frac{1}{(n+1)(n+3)} M^4 - \frac{3}{8 \pi^2} \frac{e^{-s_0^\sigma/M^2}}{(n+1)(n+3)} \left[ M^4 + s_0^s M^2 - 2 (n+3) m_q^2  M^2 \right]  \non
&&\hspace{2cm}- \langle \alpha_s G^2 \rangle \frac{1}{24 \pi} \frac{1}{n+1} - \langle {\bar q} q \rangle m_q  
+ \langle g_s {\bar q} \sigma T G q \rangle \frac{(8n+7) m_q}{18} \frac{1}{M^2} + 4\pi \alpha_s \langle {\bar q} q \rangle^2 \frac{(-4n+10)}{81} \frac{1}{M^2}.
\label{eq:zeta-sigma-even-bftsr} 
\eeq

\begin{figure}[t]
\begin{center}
\resizebox{0.45\textwidth}{!}{
\includegraphics{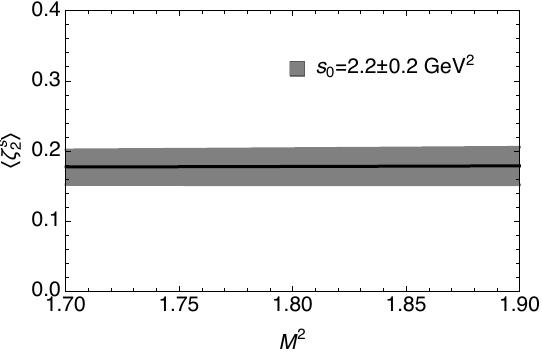}} 
\hspace{8mm}
\resizebox{0.45\textwidth}{!}{
\includegraphics{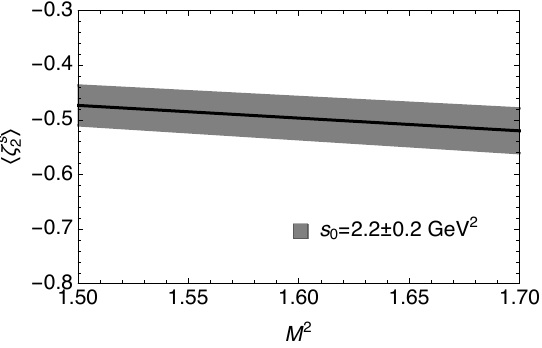}} \non
\end{center}
\vspace{-6mm}
\caption{The Borel mass dependence of the $\langle \zeta_2^{s} \rangle$ obtained from the QCDSR (left) and BFTSR (right).}
\label{fig:as2-M2}       
\end{figure}
\begin{figure}[t]
\vspace{4mm}
\begin{center}
\resizebox{0.45\textwidth}{!}{
\includegraphics{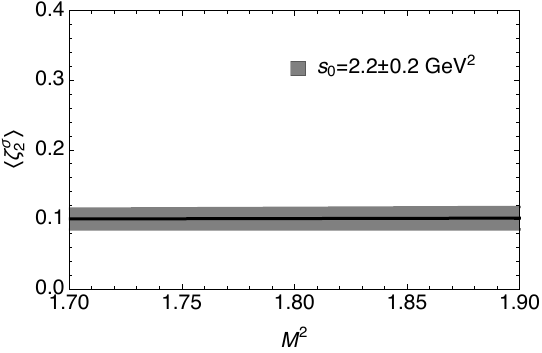}} 
\hspace{8mm}
\resizebox{0.45\textwidth}{!}{
\includegraphics{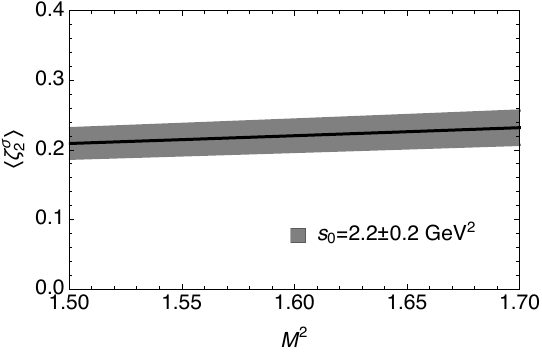}} \non
\end{center}
\vspace{-6mm}
\caption{The same as figure \ref{fig:as2-M2}, but for the second tensor moment $\langle \zeta_2^{\sigma} \rangle$.}
\label{fig:asig2-M2}       
\end{figure}

Figure \ref{fig:as2-M2} and figure \ref{fig:asig2-M2} show the Borel mass dependence of the second scalar and tensor moments. 
Its easy to find the apparent difference between the result obtained from QCDSR and BFTSR, 
and hence the different predictions for the gegenbauer coefficients as shown in table \ref{tab-f0-a1-s} and table \ref{tab-f0-a1-sig}.

\begin{table}[t]
\begin{tabular}{l |  c  c | c c  c }
\hline\hline
{\rm Sum rules} \quad & \quad {\rm QCDSR}  \quad & \quad {\rm BFTSR}  \quad & 
\quad {\rm QCDSR} \cite{Lu:2006fr} \quad & \quad {\rm BFTSR} \cite{Han:2013zg}\quad \non
\hline
$a_2^s(f_0(980))$ \quad & \quad $0.296 \pm 0.044$  \quad & \quad $-0.828 \pm 0.065 $ \quad & \quad --- \quad  &  \quad --- \quad \non
$a_2^s(f_0(1500))$ \quad & \quad --- \quad & \quad --- \quad & \quad $[-0.33, -0.18]$ \quad   &  \quad $[-0.02, 0.05]$ \quad \non
\hline
$M^2$({\rm GeV}$^2$) \quad & \quad $1.8 \pm 0.1$ \quad & \quad $1.6 \pm 0.1$ \quad & \quad $1.7 \pm 0.2$ \quad & \quad $2.0 \pm 0.2$ \quad \non
$s_0$({\rm GeV}$^2$) \quad &  \multicolumn{2}{c|}{ \quad $2.2 \pm 0.2$  \quad } & \quad $6.5 \pm 0.3$ \quad & \quad $6.5 \pm 0.3$ \quad \non
\hline\hline
\end{tabular}
\caption{The second gegenbauer coefficients obtained from the scalar sum rules.}
\label{tab-f0-a1-s}
\end{table}
\begin{table}[t]
\vspace{4mm}
\begin{tabular}{l |  c  c | c c  c }
\hline\hline
{\rm Sum rules} \quad & \quad {\rm QCDSR}  \quad & \quad {\rm BFTSR}  \quad & 
\quad {\rm QCDSR} \cite{Lu:2006fr} \quad & \quad {\rm BFTSR} \cite{Han:2013zg}\quad \non
\hline
$a_2^\sigma(f_0(980))$ \quad & \quad $0.169 \pm 0.026$  \quad & \quad $0.367 \pm 0.039$ \quad & \quad --- \quad  &  \quad --- \quad \non
$a_2^\sigma(f_0(1500))$ \quad & \quad --- \quad & \quad --- \quad  & \quad $[-0.15, -0.09]$ \quad   &  \quad $[-0.03, 0.00]$ \quad \non
\hline
$M^2$({\rm GeV}$^2$) \quad & \quad $1.8 \pm 0.1$ & \quad $1.6 \pm 0.1$ \quad & \quad $1.65 \pm 0.15$ \quad & \quad $2.0 \pm 0.2$ \quad \non
$s_0$({\rm GeV}$^2$) \quad &  \multicolumn{2}{c|}{ \quad $2.2 \pm 0.2$  \quad } & \quad $6.5 \pm 0.3$ \quad & \quad $6.5 \pm 0.3$ \quad \non
\hline\hline
\end{tabular}
\caption{The second gegenbauer coefficients obtained from the tensor sum rules.}
\label{tab-f0-a1-sig}
\end{table}

Twist three LCDAs also contributed from three particle composition \cite{Braun:2003rp,Bijnens:2002mg,Ball:2006wn}
\beq
\phi_{3S}^{q}(p_1, \alpha_i) = 360 f_{3S} \alpha_1 \alpha_2 \alpha_3^2 
\left[ 1 + \lambda_{3S} \left( \alpha_1 - \alpha_2 \right) + \omega_{3S} \left( 7 \alpha_3 - 3 \right) \right]. 
\label{eq:phi-t3-3p}
\eeq
where the nonperturbative parameters is defined by the local matrix elements
\beq
&&\langle S(p_1) \vert {\bar q}_2(z) g_s G \cdot \sigma q_1(z) \vert 0 \rangle = 2 f_{3S} \left( p_1 \cdot z \right)^2, \non
&&\langle S(p_1) \vert {\bar q}_2(z) \overleftarrow{D} g_s G \cdot \sigma q_1(z) - {\bar q}_2(z) g_s G \cdot \sigma \overrightarrow{D}  q_1(z)\vert 0 \rangle 
= - 2 f_{3S} \lambda_{3S} \left( p_1 \cdot z \right)^3/14, \non
&&\langle S(p_1) \vert {\bar q}_2(z) \sigma^{\mu\nu} \left[ i D, g_s G_{\mu\nu} \right] q_1(z) 
- \frac{3}{7} \partial {\bar q}_2(z) g_s G \cdot \sigma \overrightarrow{D}  q_1(z)\vert 0 \rangle 
= - f_{3S} \omega_{3S} \left( p_1 \cdot z \right)^3/14.
\eeq
The renormalization group equations of them are 
\beq
&&f_{3S}(\mu)= f_{3S}(\mu_0) L(\mu, \mu_0)^{-\gamma_{f_{3S}}/8}, \non
&&\left[f_{3S}\lambda_{3S}\right](\mu)= \left[f_{3S}\lambda_{3S}\right](\mu_0) L(\mu, \mu_0)^{-\gamma_{\lambda_{3S}}/8}, \non
&&\left[f_{3S}\omega_{3S}\right](\mu)= \left[f_{3S}\lambda_{3S}\right](\mu_0) L(\mu, \mu_0)^{-\gamma_{\omega_{3S}}/8}, 
\eeq
at one loop accuracy, the anomalous dimensions are 
\beq
\gamma_{f_{3S}}^{(0)} = \frac{110}{9}, \quad
\gamma_{\lambda_{3S}}^{(0)} = \frac{139}{9}, \quad
\gamma_{\omega_{3S}}^{(0)} = \frac{208}{9}.
\eeq

In the numerics we take $\lambda_{3f_0} =0$ due to the $G$-odd definition, 
and $\omega_{3f_0}({\rm 1GeV}) = -1.5 \pm 0.7$ as the same as $\omega_{3\pi}$. 
The additional scalar coupling is related to the gegengauer moments $\langle \zeta_{n=0,2}^{s/\sigma} \rangle$ and  ${\bar f}_{f_0}$ by by the equation of motion 
\beq
\langle \zeta_{2}^{s} \rangle = \frac{1}{3} \langle \zeta_{0}^{s} \rangle + \frac{4}{m_{f_0}} \frac{{\bar f}_{3f_0}}{{\bar f}_{f_0}}, \qquad 
\langle \zeta_{2}^{\sigma} \rangle = \frac{1}{5} \langle \zeta_{0}^{\sigma} \rangle + \frac{12}{5m_{f_0}} \frac{{\bar f}_{3f_0}}{{\bar f}_{f_0}} 
- \frac{8}{5m_{f_0}} \frac{{\bar f}_{3f_0}}{{\bar f}_{f_0}} \langle \langle \alpha_3 \rangle\rangle_{f_0}.
\eeq
The moments of three particle distribution amplitudes are defined 
\beq
\langle\langle \left( \alpha_2 - \alpha_1 + v \alpha_3 \right)^n \rangle\rangle = \int {\cal D} \alpha_i \phi_{3S}(\alpha_i) \left( \alpha_2 - \alpha_1 + v \alpha_3 \right)^n.
\eeq

\section{$D_s \to f_0$ form factors}\label{sec:Ds2f0-ff}

$D_s \to f_0$ transition form factors are defined by 
\beq
&& \langle f_0(p_1) \vert {\bar s} \gamma_\mu \gamma_5 c \vert D_s^+(p) \rangle 
= -i f_1(q^2) \left[ (p + p_1)_\mu - \frac{m_{D_s}^2 - m_{f_0}^2}{q^2} q_\mu \right] - i f_0(q^2) \frac{m_{D_s}^2 - m_{f_0}^2}{q^2} q_\mu \non
&&\hspace{3.5cm}  = -i \left[ f_+(q^2) \left(p + p_1 \right)_\mu + f_-(q^2) q_\mu \right]  \,. \non
&& \langle f_0(p_1) \vert {\bar s} \sigma_{\mu\nu} \gamma_5 q^\mu c \vert D_s^+(p) \rangle 
= - \frac{f_T(q^2)}{m_{D_s} + m_{f_0}} \left[ q^2 \left(p + p_1 \right)_\nu - \left( m_{D_s}^2 - m_{f_0}^2 \right) q_\nu \right] \,.
\label{eq:ff-definition}
\eeq
Here $q = p - p_1$ is the transfer momentum, the relations between two definitions are 
\beq
f_+(q^2) = f_1(q^2) \,, \quad\quad 
f_-(q^2) = \frac{m_{D_s}^2 - m_{f_0}^2}{q^2} f_0(q^2) - \frac{m_{D_s}^2 - m_{f_0}^2}{q^2} f_1(q^2) \,.
\eeq

To evaluate the $D_s \to f_0$ form factors, we start with the correlation functions 
\beq
&&\Pi_\mu^{\rm S_i}(p_1,q) = i \int d^4x e^{iqx} \langle f_0(p_1) \vert {\rm T} \{j_{1,\mu}^{\rm S_i}(x), j_{2}^{\rm S_i} \} \vert 0 \rangle,
\label{eq:correlator-v}\\
&&\tilde{\Pi}_\mu^{\rm S_i}(p_1,q) = i \int d^4x e^{iqx} \langle f_0(p_1) \vert T\{\tilde{j}_{1,\mu}^{\rm S_i}(x), j_{2}^{\rm S_i} \} \vert 0 \rangle,
\label{eq:correlator-t}
\eeq
where $j_{1,\mu}$ and $\tilde{j}_{1,\mu}$ are the weak transition currents, $j_2$ is the hadron interpolating current. 
Roman alphabets ${\rm S_i}$ denote different scenarios of the currents as shown in table \ref{tab-currents}. 
We highlight that both the leading and subleading twist LCDAs of scalar meson contribute to $D_s \to f_0$ form factors if we take the conventional non-chiral currents, 
while only the leading or subleading twist LCDAs contribute to the form factors if we take the chiral currents.

\begin{table}[tb]
\begin{center}
\begin{tabular}{c | c | c | c}
\hline\hline
Scenarios \quad & \quad $j_{1,\mu}^{\rm S_i}$ \quad\quad\quad\quad\quad $\tilde{j}_{1,\mu}^{\rm S_i}$ \quad & \quad $j_{2}^{\rm S_i}$ \quad & \quad $f_{+}, f_{-}, f_{T}$ \quad \non
\hline
${\rm S_1}$ \quad & \quad ${\bar s} \gamma_\mu \gamma_5 c$ \quad \quad \quad ${\bar s} \sigma_{\mu\nu} \gamma_5 q^\nu c$ \quad & \quad ${\bar c} i \gamma_5 s$ \quad & \quad $\phi, \phi^s, \phi^\sigma$ \quad \non
\hline
${\rm S_2}$ \quad & \quad ${\bar s} \gamma_\mu (1 - \gamma_5) c$ \quad ${\bar s} \sigma_{\mu\nu} (1 + \gamma_5) q^\nu c$ \quad & \quad ${\bar c} i (1 - \gamma_5) s$ \quad & \quad $\phi$ \quad \non
\hline
${\rm S_3}$ \quad & \quad ${\bar s} \gamma_\mu (1 - \gamma_5) c$ \quad ${\bar s} \sigma_{\mu\nu} (1 + \gamma_5) q^\nu c$ \quad & \quad ${\bar c} i (1 + \gamma_5) s$ \quad & \quad $\phi^s, \phi^\sigma$ \quad \non
\hline\hline
\end{tabular}
\vspace{-1mm}
\caption{Scenarios of the currents to evaluate the $D_s \to f_0$ form factors.}
\end{center}
\label{tab-currents}
\end{table}

In the physical region, the long-distance quark-gluon interaction between the two currents in Eqs. (\ref{eq:correlator-v},\ref{eq:correlator-t}) begins to form hadrons. 
In this respect, the correlation function can be understood by the sum of contributions from all possible intermediate states with appropriate subtractions. 
We take the (axial)-vector current in the weak vertex for example to show the dispersion relation of invariant amplitudes in variable $(p_1+q)^2 > 0$, 
which is written in 
\beq
\Pi_\mu^{\rm S_i, had}(p_1,q) = \frac{\langle f_0(p_1) \vert j_{1,\mu}^{\rm S_i}(x) \vert D_s(p_1+q) \rangle 
\langle D_s \vert j_s^{\rm S_i}(0) \vert 0 \rangle}{m_{D_s}^2 - (p_1+q)^2}  
+ \frac{1}{\pi} \int_{s^i_0}^\infty ds \frac{\rho_\mu^h(s,q^2)}{s-(p_1+q)^2}.
\label{eq:corretors-hadron}
\eeq
The ground state $D_s$ is isolated from the contributions from excited states and continuum spectra by introducing a threshold value $s_0$.
With the form factors defined in Eqs. (\ref{eq:ff-definition}) and the decay constant defined as 
$\langle D_s(p_1+q) \vert j_s^{\rm S_1}(0) \vert 0 \rangle = m_{D_s}^2 f_{D_s}/(m_c+m_s)$, 
the hadron representation is rewritten as 
\beq
\Pi_\mu^{\rm S_1, had}(p_1,q) = \frac{-i m_{D_s}^2 f_{D_s} \left[ 2 f_+(q^2) p_{1 \nu} + \left( f_+(q^2) + f_-(q^2) \right) q_\mu \right] }{ (m_c+m_s) 
\left[m_{D_s}^2 - (p_1+q)^2 \right] } + \frac{1}{\pi} \int_{s^1_0}^\infty ds \frac{\rho_+^h(s,q^2) p_{1 \mu} + \rho_-^h(s,q^2) q_{\mu} }{s-(p_1+q)^2}.
\label{eq:correlators-hadron-1}
\eeq
The relations between different scenarios read as 
\beq
\Pi_\mu^{\rm S_1, had}(p_1,q) = \Pi_\mu^{\rm S_2, had}(p_1,q) = - \Pi_\mu^{\rm S_3}(p_1,q).
\label{eq:correlators-hadron-2}
\eeq

In the Euclidean momenta space with negative $q^2$, the correlation functions can be evaluated directly by QCD at the quark-gluon level. 
Since the operator product expansion (OPE) is valid for large energies of the final state vector mesons, 
the momentum transfer squared is restriction to be not too large $0 \leqslant \vert q^2 \vert \leqslant q^2_{\rm max}$, 
and hence the operator product of the $c$-quark fields in the correlation function can be expanded near the light cone $x^2 \sim 0$ due to the large virtuality, 
\beq
&&S_c(x,0,m_c) \equiv 
- i \langle 0 \vert T\{ c_i(x), {\bar c}_j(0) \} \vert 0 \rangle = -i \frac{m_c^2}{4\pi^2} \left[ \frac{K_1(m_c\sqrt{\vert x^2 \vert})}{\sqrt{\vert x^2 \vert}} 
+ \frac{i \xsl K_2(m_c\sqrt{\vert x^2 \vert})}{\vert x^2 \vert} \right] \delta_{ij} \non
&& \hspace{1.2cm}- i \frac{g m_c}{16\pi^2} \int_0^1 du \left[ G \cdot \sigma K_0(m_c\sqrt{\vert x^2 \vert}) 
+ i \frac{ \left[ {\bar u} \xsl G \cdot \sigma +u G \cdot \sigma \xsl\right] K_1(m_c\sqrt{\vert x^2 \vert})}{\sqrt{\vert x^2 \vert}} \right] \delta_{ij}+ \cdots. 
\label{eq:c-propagator}
\eeq
The first term corresponds to the free charm quark propagator, the second one corresponds to the quark-gluon interaction at leading power, 
and the ellipsis denotes the high power corrections from the quark-gluon interaction. ${\bar u} = 1 - u$ is indicated. 
The correlation functions are ultimately written in a general convolution of hard functions with various LCDAs at different twists
\beq
\Pi_\mu^{\rm S_i, OPE}(p_1,q) = \sum_{t} \int_0^1 du \, T^{(t)}_\mu(u, q^2, (p_1+q)^2) \otimes \phi^{(t)}(u) 
+ \int_0^1 du \int_0^u {\cal D} \alpha_i  \, T^\prime_\mu(u, \alpha_i, q^2, (p_1+q)^2) \otimes \phi_{3f_0}(\alpha_i), 
\label{eq:correlator-qcd}
\eeq
where the first term comes from the two particle LCDAs, and the second term comes from the twist three LCDA with three particle configuration. 
The OPE amplitudes in Eq. (\ref{eq:correlator-qcd}) can also be written in a dispersion integral over the invariant mass of the interpolating heavy meson
\beq
\Pi_\mu^{\rm S_i, OPE}(p_1,q) &=& \frac{1}{\pi} \int_0^1 du \, \sum_{n =1, 2} 
\left[ \frac{{\rm Im} \Pi_{+,n}^{\rm S_i, OPE}(q^2, u) \, p_{1 \mu}+ {\rm Im} \Pi_{-,n}^{\rm S_i, OPE}(q^2, u) q_\mu}{u^n \left[s_2(u) - \left(p_1+q\right)^2\right]^n} \right.\non 
&& \left. + \int_0^u {\cal D}\alpha_i \frac{{\rm Im} \Pi_{+,n}^{\rm S_i, OPE}(q^2, u,\alpha_i) \, p_{1 \mu} 
+ {\rm Im} \Pi_{-,n}^{\rm S_i, OPE}(q^2, u,\alpha_i) \, q_\mu}{\left[\alpha_2 + v \left(1-\alpha_1-\alpha_2 \right)\right]^n \left[s_3(u,\alpha_i) - \left(p_1+q\right)^2\right]^n} \right]. 
\label{eq:DR-OPE}
\eeq
Here the momentum fraction dependent kinematical variables read as $s_2(u) = {\bar u} m_{f_0}^2 + (m_c^2 - {\bar u} q^2)/u$ and 
$s_3(u,\alpha_i) = (1 - \alpha_2 - v \alpha_3) m_{f_0}^2 + [m_c^2 - (1 - \alpha_2 - v \alpha_3) q^2 ]/(\alpha_2 + v \alpha_3)$. 

We then implement the quark-hadron duality to eliminate the contribution from the excited and continuum spectra 
with the threshold invariant mass $s_0^{i}$. 
In order to improve the reliability of quark-hadron duality, we Borel transfer both the hadronic representation and the OPE evaluation of the correlation functions. 
This operation, from one hand, sticks out the ground scalar meson by suppressing the contribution from the excited and continuum spectra in the hadron representation, 
from the other hand, is helpful to obtain the convergent power expansion in the OPE evaluation. 
The $D_s \to f_0$ form factors obtained from LCSRs approach with different currents are collected as follow, 
\beq
&~&f_+^{\rm S1}(q^2) = \frac{m_c + m_s}{2m_{D_s}^2 f_{D_s}} \left\{ 
\int_{u_0}^1 \frac{du}{u} \left[ -m_c \phi(u) + u m_{f_0} \phi^s(u) + \frac{m_{f_0} \phi^\sigma(u)}{3} 
+ \frac{m_{f_0} \phi^\sigma(u)}{6} \frac{m_c^2 + q^2 - u^2m_{f_0}^2}{uM^2} \right] e^{\frac{-s_2(u) + m_{D_s}^2}{M^2}} \right. \non
&~& \hspace{1.2cm} + \left. \int_{u_0}^1 du \int_0^u d \alpha_1 \int_0^{1-u} \frac{d \alpha_2}{\alpha_3} 
\frac{8 m_{f_0}^2 f_{3f_0}^s \phi_{3f_0}^s(\alpha_i)}{\left[\alpha_2 + v \alpha_3 \right] M^2} \, e^{\frac{-s_3(u,\alpha_i) + m_{D_s}^2}{M^2}} 
+\frac{m_{f_0} \phi^\sigma(u_0)}{6} \frac{m_c^2 + q^2 - u_0^2m_{f_0}^2}{m_c^2 - q^2 + u_0^2m_{f_0}^2} \, e^{\frac{-s_0^1 + m_{D_s}^2}{M^2}} \right. \non
&~& \hspace{1.2cm} +  \left.  \int_0^{u_0} d \alpha_1 \int_0^{1-u_0} \frac{d \alpha_2}{\alpha_3} 
\frac{8 m_{f_0}^2 f_{3f_0}^s \phi_{3f_0}^s(\alpha_i) \, \left(\alpha_2 + v_0 \alpha_3 \right)}{\left[m_c^2 - q^2 + \left( \alpha_2 + v_0 \alpha_3 \right)^2 m_{f_0}^2\right]} \, 
e^{\frac{-s_0^1 + m_{D_s}^2}{M^2}} \right\}, \\
\label{f+S1} 
&~&f_+^{\rm S1}(q^2) + f_-^{\rm S1}(q^2) = \frac{m_c + m_s}{m_{D_s}^2 f_{D_s}} \left\{ 
\int_{u_0}^1 \frac{du}{u} \left[ m_{f_0} \phi^s(u) + \frac{m_{f_0} \phi^\sigma(u)}{6u} 
- \frac{m_{f_0} \phi^\sigma(u)}{6} \frac{m_c^2 - q^2 + u^2m_{f_0}^2}{u^2M^2} \right] e^{\frac{-s_2(u) + m_{D_s}^2}{M^2}} \right. \non
&~& \hspace{1.2cm} + \left. \int_{u_0}^1 du \int_0^u d \alpha_1 \int_0^{1-u} \frac{d \alpha_2}{\alpha_3} 
\frac{8 m_{f_0}^2 f_{3f_0}^s \phi_{3f_0}^s(\alpha_i)}{\left[\alpha_2 + v \alpha_3 \right]^2 M^2} \, e^{\frac{-s_3(u,\alpha_i) + m_{D_s}^2}{M^2}} 
- \frac{m_{f_0} \phi^\sigma(u_0)}{6 u_0} \, e^{\frac{-s_0^1 + m_{D_s}^2}{M^2}} \right. \non
&~& \hspace{1.2cm} + \left.  \int_0^{u_0} d \alpha_1 \int_0^{1-u_0} \frac{d \alpha_2}{\alpha_3} 
\frac{8 m_{f_0}^2 f_{3f_0}^s \phi_{3f_0}^s(\alpha_i)}{\left[m_c^2 - q^2 + \left( \alpha_2 + v_0 \alpha_3 \right)^2 m_{f_0}^2 \right]} \, 
e^{\frac{-s_0^1 + m_{D_s}^2}{M^2}} \right\}, \\
\label{f-S1} 
&~&f_T^{\rm S1}(q^2) = \frac{\left(m_c + m_s\right)\left(m_{D_s}+m_{f_0}\right)}{m_{D_s}^2 f_{D_s}} \left\{ 
\int_{u_0}^1 \frac{du}{u} \left[ - \frac{\phi(u)}{2} + \frac{m_{f_0} \phi^\sigma(u)}{6}\frac{m_c}{u M^2} \right] e^{\frac{-s_2(u)+m_{D_s}^2}{M^2}} \right. \non
&~& \hspace{1.2cm} + \left. \frac{m_{f_0} \phi^\sigma(u_0)}{6} \frac{m_c}{\left[ m_c^2 - q^2 + u_0^2m_{f_0}^2 \right]} \, e^{\frac{-s_0^1+m_{D_s}^2}{M^2}}  \right\}, 
\label{fTS1} \\
&~&f_+^{\rm S2}(q^2) = - \frac{m_c \left(m_c + m_s\right)}{m_{D_s}^2 f_{D_s}} \int_{u_0}^1 \frac{du}{u} \phi(u) e^{\frac{-s_2(u) + m_{D_s}^2}{M^2}} , \\
\label{f+S2} 
&~&f_+^{\rm S2}(q^2) + f_-^{\rm S2}(q^2) = 0, \\
\label{f-S2} 
&~&f_T^{\rm S2}(q^2) = - \frac{\left(m_c + m_s\right)\left(m_{D_s}+m_{f_0}\right)}{m_{D_s}^2 f_{D_s}} 
\int_{u_0}^1 \frac{du}{u} \phi(u) e^{\frac{-s_2(u)+m_{D_s}^2}{M^2}} , 
\label{fTS2} \\
&~&f_+^{\rm S3}(q^2) = \frac{\left(m_c + m_s\right) m_{f_0}}{2 m_{D_s}^2 f_{D_s}} \left\{ 
\int_{u_0}^1 \frac{du}{u} \left[ 2 u  \phi^s(u) + \frac{2 \phi^\sigma(u)}{3} 
+ \frac{\phi^\sigma(u)}{3} \frac{m_c^2 + q^2 - u^2m_{f_0}^2}{uM^2} \right] e^{\frac{-s_2(u) + m_{D_s}^2}{M^2}} \right. \non
&~& \hspace{1.2cm} + \left. \int_{u_0}^1 du \int_0^u d \alpha_1 \int_0^{1-u} \frac{d \alpha_2}{\alpha_3} 
\frac{16 m_{f_0} f_{3f_0}^s \phi_{3f_0}^s(\alpha_i)}{\left[\alpha_2 + v \alpha_3 \right] M^2} \, e^{\frac{-s_3(u,\alpha_i) + m_{D_s}^2}{M^2}} 
+\frac{\phi^\sigma(u_0)}{3} \frac{m_c^2 + q^2 - u_0^2m_{f_0}^2}{m_c^2 - q^2 + u_0^2m_{f_0}^2} \, e^{\frac{-s_0^1 + m_{D_s}^2}{M^2}} \right. \non
&~& \hspace{1.2cm} + \left.  \int_0^{u_0} d \alpha_1 \int_0^{1-u_0} \frac{d \alpha_2}{\alpha_3} 
\frac{16 m_{f_0} f_{3f_0}^s \phi_{3f_0}^s(\alpha_i) \left(\alpha_2 + v_0 \alpha_3 \right)}{\left[m_c^2 - q^2 + \left( \alpha_2 + v_0 \alpha_3 \right)^2 m_{f_0}^2\right]} \, 
e^{\frac{-s_0^1 + m_{D_s}^2}{M^2}} \right\}, 
\label{f+S3} \\
&~&f_+^{\rm S3}(q^2) + f_-^{\rm S3}(q^2) = \frac{\left(m_c + m_s\right) m_{f_0}}{m_{D_s}^2 f_{D_s}} \left\{ 
\int_{u_0}^1 \frac{du}{u} \left[ 2 \phi^s(u) + \frac{\phi^\sigma(u)}{3u} 
- \frac{\phi^\sigma(u)}{3} \frac{m_c^2 - q^2 + u^2m_{f_0}^2}{u^2M^2} \right] e^{\frac{-s_2(u) + m_{D_s}^2}{M^2}} \right. \non
&~& \hspace{1.2cm} + \left. \int_{u_0}^1 du \int_0^u d \alpha_1 \int_0^{1-u} \frac{d \alpha_2}{\alpha_3} 
\frac{16 m_{f_0} f_{3f_0}^s \phi_{3f_0}^s(\alpha_i)}{\left[\alpha_2 + v \alpha_3 \right]^2 M^2} \, e^{\frac{-s_3(u,\alpha_i) + m_{D_s}^2}{M^2}} 
- \frac{\phi^\sigma(u_0)}{3 u_0} \, e^{\frac{-s_0^1 + m_{D_s}^2}{M^2}} \right. \non
&~& \hspace{1.2cm} + \left.  \int_0^{u_0} d \alpha_1 \int_0^{1-u_0} \frac{d \alpha_2}{\alpha_3} 
\frac{16 m_{f_0} f_{3f_0}^s \phi_{3f_0}^s(\alpha_i)}{\left[m_c^2 - q^2 + \left( \alpha_2 + v_0 \alpha_3 \right)^2 m_{f_0}^2 \right]} \, e^{\frac{-s_0^1 + m_{D_s}^2}{M^2}} \right\}, 
\label{f-S3} \\
&~&f_T^{\rm S3}(q^2) = \frac{\left(m_c + m_s\right)\left(m_{D_s}+m_{f_0}\right) m_c m_{f_0}}{m_{D_s}^2 f_{D_s}} \left\{ 
\int_{u_0}^1 \frac{du}{u} \frac{\phi^\sigma(u)}{3 M^2} \, e^{\frac{-s_2(u)+m_{D_s}^2}{M^2}} 
+ \frac{\phi^\sigma(u_0)}{3 \left[ m_c^2 - q^2 + u_0^2m_{f_0}^2 \right]} \, e^{\frac{-s_0^1+m_{D_s}^2}{M^2}}  \right\}. 
\label{fTS3} 
\eeq
The threshold momentum fraction is the solution of $s_i(u_0) = s_0$ with $i=1,2$, 
\beq
u_0^{i} = \frac{- (s_0^i - q^2 - m_{f_0}^2) + \sqrt{(s_0^i - q^2 - m_{f_0}^2)^2 + 4 m_{f_0}^2 (m_c^2 - q^2)}}{2 m_{f_0}^2}.
\eeq

The expressions in Eqs. (\ref{f+S1}-\ref{fTS1}), Eqs. (\ref{f+S2}-\ref{fTS2}) and Eqs. (\ref{f+S3}-\ref{fTS3}), under different scenarios of the currents, 
are consistent with the result obtained in Ref. \cite{Colangelo:2010bg}, Ref. \cite{Sun:2010nv} and Ref. \cite{Han:2013zg}, respectively. 
The above calculations are based on the ideal two-quark configuration that $f_0$ is pure an $s{\bar s}$ state. 
Nevertheless, there are several experiment measurements indicate the mixing between $f_0$ and $\sigma$, 
\beq
\sigma = \vert {\bar n}n \rangle \cos \theta + \vert {\bar s}s \rangle \sin \theta, \qquad
f_0 = - \vert {\bar n}n \rangle \sin \theta + \vert {\bar s}s \rangle \cos \theta.
\eeq
Then the form factor in Eqs. (\ref{f+S1}-\ref{fTS1}), Eqs. (\ref{f+S2}-\ref{fTS2}) and Eqs. (\ref{f+S3}-\ref{fTS3}) are modified\footnote{The mixing would not bring changes to the LCSRs in section \ref{sec:LCDAs} since the angle dependence could be absorbed in to the definition of scalar decay constant.} by multiplying an angle dependence $\cos \theta$. 
The mixing angle extracted from the data is not larger than $40\degree$ \cite{Gokalp:2004ny,Fleischer:2011au,Cheng:2022vbw},
and a recent LHCb measurement of the upper limit on the branching fraction ${\cal B}({\bar B}^0 \to J/\Psi f_0) \times {\cal  B}(f_0 \to \pi^+\pi^-)$ 
leads to $\vert \theta \vert < 30\degree$ \cite{LHCb:2013dkk}. 
In the follow calculation, we would take the mixing angle $\theta = 20\degree \pm 10\degree$. 
		 
The value of Borel mass squared is implied by the internal virtuality of propagator which is smaller than the cutoff threshold value, 
saying $M^2 \sim \mathcal{O}(u m_{D_s}^2 + {\bar u} q^2 - u {\bar u} m_{f_0}^2) < s_0$, 
this value is a litter bit larger than the factorisation scale we chosen at $\mu_f^2 = m^2_{D_s} - m_c^2 = 1.48^2 \, {\rm GeV}^2$ 
with the quark mass ${\overline m_c}(m_c) = 1.30 \, {\rm GeV}$. 
In practice the selection of Borel mass is actually a compromise between the 
overwhelming chosen of ground state in hadron spectral that demands a small value 
and the convergence of OPE evaluation that prefers a large one, 
which result in a region where $H_{ij}(q^2)$ shows an extremum in $M^2$
\beq 
\frac{d}{d(1/M^2)} {\rm ln} H_{ij}(q^2) = 0 \,.
\eeq
The continuum threshold is usually set to close to the outset of the first excited state with the same quantum number as $D_s$ 
and characterised by $s_0 \approx (m_{D_s} + \chi)^2$, which is finally determined by considering the maximal stable
evolution of physical quantities on the Borel mass squared. We take $s_0 \equiv s_0^1 = s_0^2$ in the numerics. 
The chose of these two parameters should guaratee the convergence of twist expansion in the truncated OPE calculation 
(high twists contributions are no more than thirty percents) and simultaneously the high energy cutoff in the hadron interpolating 
(the contributions from high excited state and continuum spectral is smaller than thirty percents). 
We finally set them at $M^2 = 5.0 \pm 0.5$ GeV$^2$ and $s_0 = 6.0 \pm 0.5$ GeV$^2$ in this work. 
The value of Borel mass is a litter bit larger than it chosen in the $D_s \to \pi, K$ transition\cite{Khodjamirian:2000ds}, 
close to it chosen in the $D_s \to \phi$ \cite{Colangelo:2010bg}, $D_s \to \eta^{\prime}$ \cite{Offen:2013nma} 
and $D_s \to f_0(980)$ transition \cite{Bediaga:2003hr}. 

We show the LCSRs result of $D_s \to f_0(980)$ form factors in table \ref{Ds2f0-ff-currents}, 
the results obtained from other approaches are also presented parallel for comparison. 
We see that the result obtained by adopting different currents (${\rm S_1,S_2,S_3}$) are different, 
especially for the form factors $f_+$ which gives the contribution to semileptonic $D_s^+ \to f_0 e^+ \nu$ decays. 
The difference can be traced back to the ill-defined sum rules with the chiral currents under scenario ${\rm S_2}$ and ${\rm S_3}$, 
in which only the axial-vector current ${\bar s} \gamma_\mu \gamma_5 c$ is considered in the $D_s \to f_0$ decay 
while the vector current ${\bar s} \gamma_\mu c$ with the $D_{s0}^\ast \to f_0$ decay is overlooked at the hadron level. 
Hereafter we would pay attention on the sum rules with the current chosen under scenario ${\rm S_1}$. 
Our result $f_+(0) = 0.58 \pm 0.07$ is much larger than the previous LCSRs calculation $f_+(0)=0.30 \pm 0.03$ \cite{Colangelo:2010bg},
but more consistent with the recent measurement $0.52 \pm 0.05$ at BESIII \cite{BESIII:2023wgr}.  
The main reason of the difference is the input of decay constant ${\tilde f}_{f_0}$, 
which is taken at $180$ MeV \cite{DeFazio:2001uc} in the previous work but $335$ MeV evaluated from QCDSR here. 
Besides it, we have added the first gegenbauer expansion terms in the LCDAs which part contributions are ignored in the previous work.

\begin{table}[t]
\begin{tabular}{c | r | r | r | c | c | c | c | c}
\hline\hline
{\rm FFs} \quad & \quad LCSRs-${\rm S_1}$ \quad & \quad LCSRs-${\rm S_2}$ \quad & \quad LCSRs-${\rm S_3}$ \quad & 3pSRs\cite{Bediaga:2003hr} & 3pSRs\cite{Aliev:2007uu} & LFQM\cite{Ke:2009ed} & CLFD/DR\cite{El-Bennich:2008rkp} & LCSRs\cite{Colangelo:2010bg} \non
\hline
$f_+(0)$ \quad & $0.58 \pm 0.07$ \quad & $0.40 \pm 0.06$ \quad & $0.78^{+0.13}_{-0.10}$ \quad & $0.80 \pm 0.08$ & $0.96$ & $0.87$ & $0.86/0.90$ & $0.30 \pm 0.03$  \non
\hline
$f_-(0)$ \quad & $-0.10^{-0.04}_{+0.03}$ \quad & $-0.40 \mp 0.06$ \quad & $0.21 \pm 0.04$ \quad & \quad --- \quad & \quad --- \quad & \quad --- \quad & \quad $0$ \quad & \quad --- \quad  \non
\hline
$f_T(0)$ \quad & $0.79 \pm 0.13$ \quad & $0.95 \pm 0.16$ \quad & $0.47^{+0.07}_{-0.06}$ \quad& \quad --- \quad & \quad --- \quad & \quad --- \quad & \quad --- \quad & \quad --- \quad   \non
\hline\hline
\end{tabular}
\vspace{-1mm}
\caption{$D_s \to f_0$ form factors at the full recoiled point obtained from various theoretical work under the ${\bar s}s$ description of $f_0$.}
\label{Ds2f0-ff-currents}
\end{table}

\begin{figure}[th]
\resizebox{0.3\textwidth}{!}{
\includegraphics{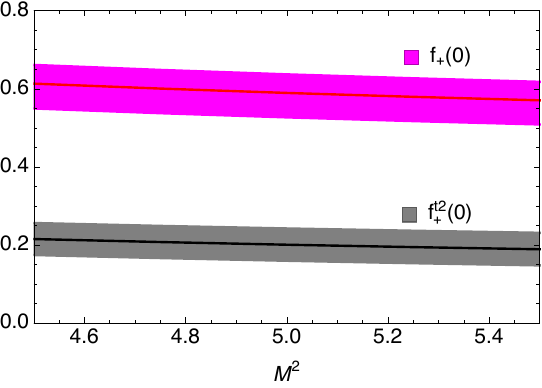}} 
\hspace{4mm}
\resizebox{0.3\textwidth}{!}{
\includegraphics{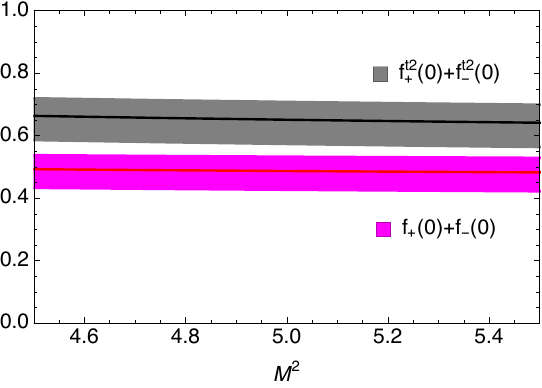}} 
\hspace{4mm}
\resizebox{0.3\textwidth}{!}{
\includegraphics{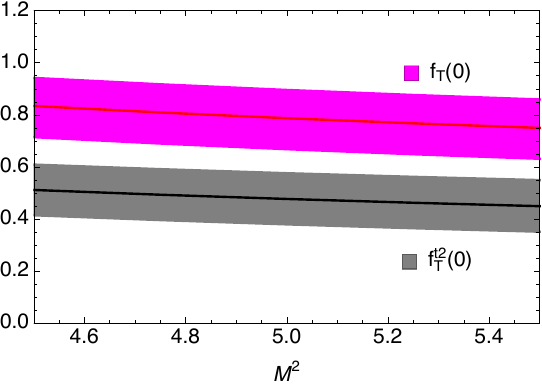}} 
\non
\vspace{-2mm}
\caption{The Borel mass dependence of $D_s \to f_0$ form factors from the LCSRs. }
\label{fig:Ds2f0-M2}       
\end{figure}
\begin{figure}[th]
\resizebox{0.32\textwidth}{!}{
\includegraphics{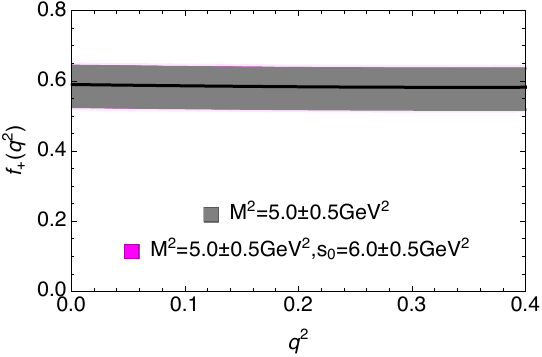}} 
\hspace{1mm}
\resizebox{0.32\textwidth}{!}{
\includegraphics{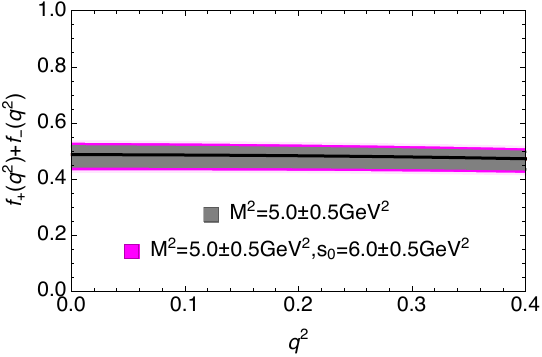}} 
\hspace{1mm}
\resizebox{0.32\textwidth}{!}{
\includegraphics{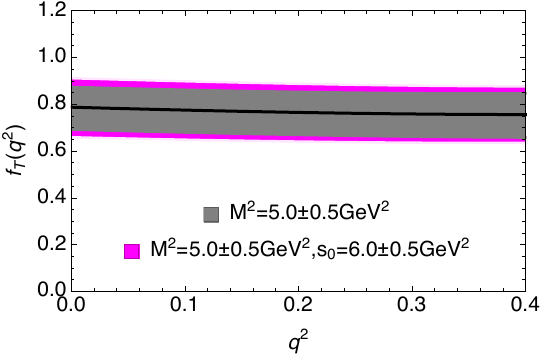}} 
\non
\vspace{-2mm}
\caption{The $q^2$ dependence of $D_s \to f_0$ form factors from the LCSRs.}
\label{fig:Ds2f0-q2}       
\end{figure}

We show the dependence of form factors on the Borel mass in figure \ref{fig:Ds2f0-M2}, 
where the gray and magnate curves correspond to the result obtained up to leading twist and subleading twist LCDAs, 
the uncertainties come from the threshold value $s_0$.
We see that the subleading twist contribution is dominate in the form factors $f_+$ , 
while the leading twist contribution become more important for the form factor $f_-$ and $f_T$. 
We plot the $q^2$ dependence of the form factors in figure \ref{fig:Ds2f0-q2} with the LCSRs maximal momentum transfer 
$q^2_{\rm max} = 0.4 \, {\rm GeV}^2$ \cite{Cheng:2022mvd}, 
where the uncertainties associated to the Borel mass $M^2$ and also the threshold value $s_0$ are shown in the gray and magnate bands, respectively.
We find that the uncertaintes of LCSRs prediction to $D_s \to f_0$ form factors is full dominated by $M^2$. 
To estimate the uncertainties associated to scale $\mu_f$, we vary the charm quark mass in the region ${\bar m_c}(m_c) = 1.30 \pm 0.30 \, {\rm GeV}$ 
which result in $\mu_f = 1.48 \pm 0.30 \, {\rm GeV}$, 
we find this variation bring another $20\%$ uncertainty to $f_+(q^2)$ and $f_T(q^2)$, and bring much significant corrections to $f_-(q^2)$ (large than $50\%$).

\subsection{$B_s \to f_0$ form factors}\label{ssec:Bs2f0-ff}

\begin{table}[tb]
\begin{tabular}{c | r | r | r | c | c | c | c | c}
\hline\hline
{\rm FFs} \quad & \quad LCSRs-${\rm S_1}$ \quad & \quad LCSRs-${\rm S_2}$ \quad & \quad LCSRs-${\rm S_3}$ \quad & PQCD\cite{Li:2008tk} & QCDSR\cite{Ghahramany:2009zz} & LCSR-chiral\cite{Sun:2010nv} & LCSRs\cite{Colangelo:2010bg} & LCSRs\cite{Cheng:2019tgh} \non
\hline
$f_+(0)$ \quad & $0.45^{+0.09}_{-0.07}$ \quad & $0.71 \pm 0.17$ \quad & $0.19 \pm 0.02$ \quad & $0.70$ & $0.12$ & $0.44$ & $0.37$ & $0.52$  \non
\hline
$f_-(0)$ \quad & $-0.41^{-0.11}_{+0.12}$ \quad & $-0.71\mp 0.17$ \quad & $-0.13 \mp 0.03$ \quad & $0$  & ---  & $-0.44$ & $0$ &  $0.04$  \non
\hline
$f_T(0)$ \quad & $0.54^{+0.09}_{-0.12}$ \quad & $0.89^{+0.20}_{-0.22}$ \quad & $0.16 \pm 0.02$ \quad& $0.40$ & $-0.08$ & $0.58$ & $0.228$ & $0.21$  \non
\hline\hline
\end{tabular}
\vspace{-1mm}
\caption{$B_s \to f_0$ form factors at the full recoiled point obtained from various theoretical work under the ${\bar s}s$ description of $f_0$.}
\label{Bs2f0-ff-currents}
\end{table}

With the definitions of heavy to light transition form factors in Eq. (\ref{eq:ff-definition}), 
we can straightly obtain the $B_s \to f_0$ form factor by substituting the charm quark to bottom quark in the LCSRs result shown in Eqs. (\ref{f+S1}-\ref{fTS1}). 
We take the bottom quark mass at ${\bar m_b}(m_b) = 4.2 \, {\rm GeV}$ 
and the factorization scale at $\mu_f = {\cal O}(m_{B_s}^2 - m_b) = 3$ GeV following the work in Ref. \cite{Duplancic:2008ix}. 
The Borel mass and threshold value are chosen at $M^2 = 18 \pm 2 \, {\rm GeV}^2$ and $s_0 = 36 \pm 2\, {\rm GeV}^2$ 
closing to the choice in the previous LCSRs work \cite{Colangelo:2010bg}. 
In table \ref{Bs2f0-ff-currents}, we present the $B_s \to f_0$ form factors at the full recoiled point obtained from various theoretical approaches, 
like the perturbative QCD(PQCD) \cite{Li:2008tk}, QCDSR \cite{Ghahramany:2009zz}, LCSRs with chiral current \cite{Sun:2010nv}, 
LCSRs with light meson on-shell \cite{Colangelo:2010bg}, LCSRs with $B$ meson on-shell \cite{Cheng:2019tgh}. 
Again, our result with different weak vertex currents (${\rm S_1,S_2,S_3}$) have large difference 
and we focus on the result obtained under scenario ${\rm S_1}$ with the axial-vector weak current. 
Our result is very close to the previous LCSRs result with the chiral current \cite{Sun:2010nv} which we would like to consider it as an incident. 
With using the same LCDAs of $f_0$, our result is different from the result obtained in previous LCSRs \cite{Colangelo:2010bg},  
this is largely because of the different input of scalar coupling ${\tilde f}_{f_0}$ too. 
We can see also that our result of $f_+(0)$ is very close to it obtained in the LCSRs with the $B$ meson LCDAs \cite{Cheng:2019tgh}, 
while the form factors $f_-(0)$ and $f_T(0)$ have large deviations, so the high order corrections are highly eager to improve the accuracy.
Here we estimate the possible NLO correction by varying the factoriazation scale as $\mu_f = 3.0 \pm 0.5$ GeV, 
which in turn brings another $20\%-30\%$ uncertainty to $f_+(q^2)$, $f_-(q^2)$ and $f_T(q^2)$.

\begin{figure}[t]
\resizebox{0.3\textwidth}{!}{
\includegraphics{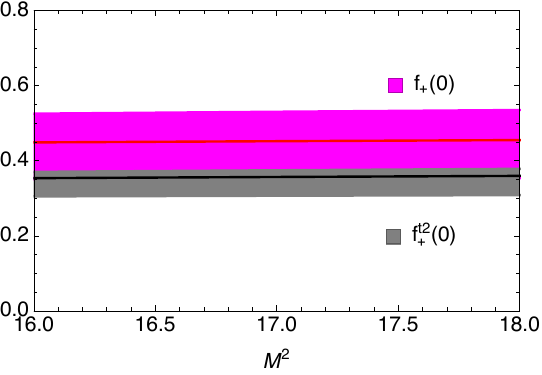}} 
\hspace{4mm}
\resizebox{0.3\textwidth}{!}{
\includegraphics{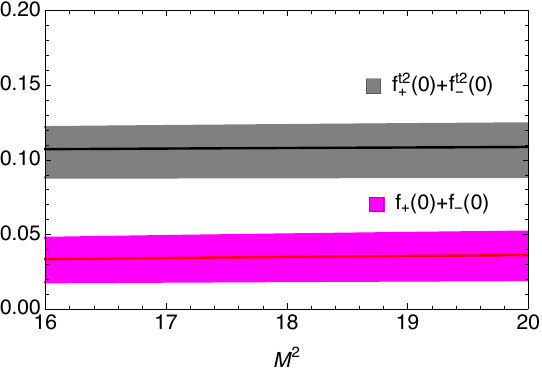}} 
\hspace{4mm}
\resizebox{0.3\textwidth}{!}{
\includegraphics{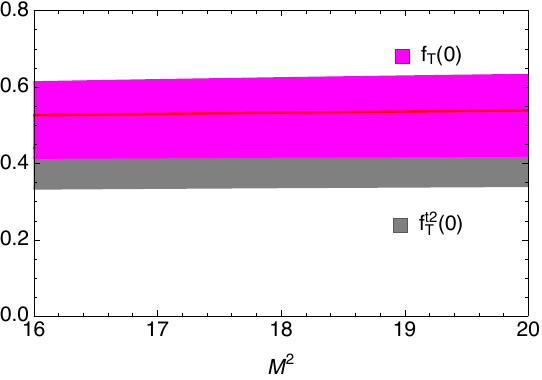}} 
\non
\vspace{-2mm}
\caption{The Borel mass dependence of $B_s \to f_0$ form factors from the LCSRs. }
\label{fig:Bs2f0-M2}       
\end{figure}
\begin{figure}[t]
\resizebox{0.32\textwidth}{!}{
\includegraphics{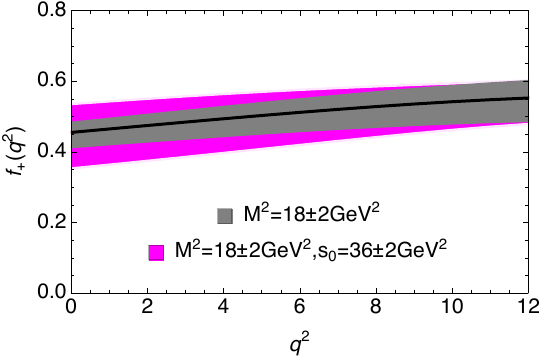}} 
\hspace{1mm}
\resizebox{0.32\textwidth}{!}{
\includegraphics{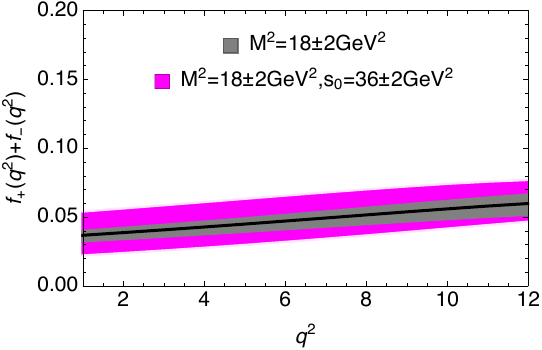}} 
\hspace{1mm}
\resizebox{0.32\textwidth}{!}{
\includegraphics{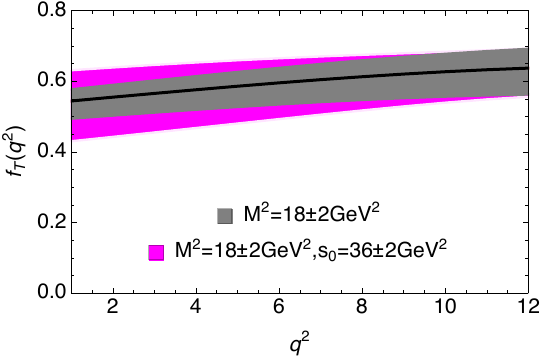}} 
\non
\vspace{-2mm}
\caption{The $q^2$ dependence of $B_s \to f_0$ form factors from the LCSRs.}
\label{fig:Bs2f0-q2}       
\end{figure}

In figure \ref{fig:Bs2f0-M2} we depict the Borel mass dependence of $B_s \to f_0$ form factors where the LCSRs result with the accuracy up to 
leading twist and subleading twist LCDAs are shown in gray and magnate curves, respectively. 
Due to the much better heavy quark expansion convergence in contrast to the $D_s$ decay, 
we see that the leading twist contribution is the dominate one in $B_s \to f_0$ transition. 
We plot the momentum transfer dependence in figure \ref{fig:Bs2f0-q2} with the largest momentum transfer at $q_{\rm max}^2 = 12 \,{\rm GeV}^2$,
where the uncertainties associated to Bore mass and also threshold value are overlap drawing in gray and magnate curves. 
We see that the dominate source of uncertainty comes from both the them. 

\subsection{$D_s^+ \to (f_0 \to ) \left[ \pi\pi \right]_{\rm S} e^+ \nu_e$ decay}\label{sec:Ds2f0-pipi}

The second-order differential decay width of semileptonic decays $D_s^+(p) \to f_0(p_1) e^+ (p_2) \nu_e(p_3)$ is proportional to the transition form factor $f_1$ as 
\beq
\frac{d^2\Gamma(D_s^+ \to f_0 e^+ \nu_e)}{dE_2 dq^2} = \frac{G_F^2 m_{D_s}^2 \vert V_{cs}\vert^2}{16 \pi^3}  \vert f_1(q^2) \vert^2 \left[ 2x(1+y-z) - 4x^2 -y \right], 
\eeq
in which the dimensionless quantities are
\beq
x \equiv \frac{E_2}{m_{D_s}} \, \quad y \equiv \frac{q^2}{m_{D_s^\ast}^2} \,, \quad z  \equiv \frac{m_{f_0}^2}{m_{D_s}^2} 
\eeq
with $E_2$ being the energy of lepton and $q^2 \equiv m_{23}^2 = (p_2 + p_3)^2 = (p - p_1)^2$ being the invariant mass of lepton-neutrino pair. 
Integrating over the lepton energy, we obtain the differential decay width on the momentum transfers $q^2$
\beq
\frac{d\Gamma(D_s^+ \to f_0 e^+ \nu_e)}{dq^2} = \frac{G_F^2  \vert V_{cs}\vert^2  \lambda^{3/2}(m_{D_s}^2, m_{f_0}^2, q^2)}{192 \pi^3 m_{D_s}^3} \vert f_+(q^2) \vert^2.
\label{eq:decaywidth-massless}
\eeq
$\lambda(a,b,c) = a^2 + b^2 + c^2 - 2xy - 2xz - 2yz$ is the Kall\"en function. 

From the experimental side, $f_0$ is measured via the $\left[ \pi\pi \right]_{\rm S}$ invariant mass spectral. 
So the key question in the phenomena is to look close at the role of $f_0$ in the $\pi\pi$ invariant mass. 
A natural solution is to examine the effects of resonance width, 
and noticeability, $D_s(p) \to \left[ \pi(k_1)\pi(k_2) \right]_{\rm S} l(p_2) \nu(p_3)$ decay is the kinematically simplest channel, 
because the invariant amplitude depends only on the invariant mass of dipion system $s \equiv p_1^2  = (k_1 + k_2)^2$ 
and not rely on its angular orientation with respecting the the remaining particles, 
meanwhile it is the most important channel due to the $\left[ \pi\pi \right]_{\rm S}$ phase shift shows a very board rise. 

To perform with the BESIII analysis \cite{BESIII:2023wgr}, we take the Flatt\'e model to describe the width effect of intermediate resonant. 
The second-order differential decay width of $D_s^+ \to (f_0 \to ) \left[ \pi\pi \right]_{\rm S} e^+ \nu_e$ is then written in 
\beq
\frac{d^2\Gamma(D_s^+ \to \left[ \pi\pi \right]_{\rm S} e^+ \nu_e)}{dE_2 dq^2} 
= \frac{1}{\pi} \int_{4m_\pi^2}^{s_{\rm max}} ds \frac{g_1^2 \beta_{\pi\pi}}{ \vert s - m_{\rm S}^2 + i (g_1^2 \beta_{\pi\pi} +g_2^2 \beta_{KK}) \vert^2} 
\frac{d^2\Gamma(D_s^+ \to {\rm S} l^+ \nu)}{dE_2 dq^2}.
\eeq
$\beta_{\pi\pi}(s) = \sqrt{1-4m_\pi^2/s}$ and $\beta_{KK}(s) = \sqrt{1-4m_K^2/s}$ are the dipion phase factors, 
$g_1^2 = 0.165$ GeV$^2$ and $g_2^2 = 0.695$ GeV$^2$ are the weighted parameters \cite{BES:2004twe}. 
Integrating over the invariant mass one arrives at the differential decay width on the momentum transfers 
\beq
\frac{d\Gamma(D_s^+ \to \left[ \pi\pi \right]_{\rm S} e^+ \nu_e)}{dq^2} = 
\frac{1}{\pi} \frac{G_F^2 \vert V_{cs}\vert^2}{192 \pi^3 m_{D_s}^3} \vert f_+(q^2) \vert^2
\int_{4m_\pi^2}^{s_{\rm max}} ds \frac{\lambda^{3/2}(m_{D_s}^2, s, q^2) \, g_1^2 \beta_{\pi\pi}(s)}{ \vert s - m_{\rm S}^2 + i (g_1^2 \beta_{\pi\pi}(s) +g_2^2 \beta_{KK}(s)) \vert^2}.
\label{eq:decaywidth-resonant}
\eeq

\begin{figure}[tb]
\begin{center}
\resizebox{0.45\textwidth}{!}{
\includegraphics{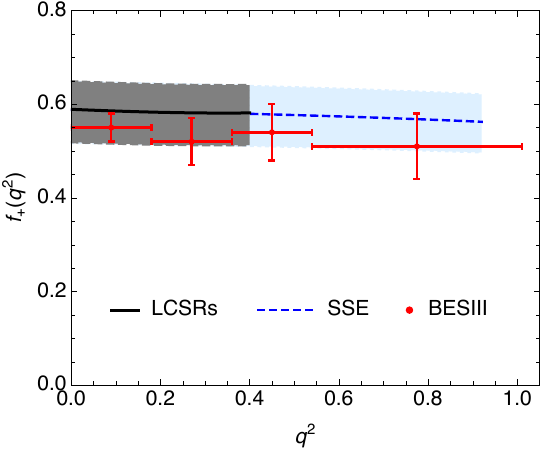}} 
\hspace{8mm}
\resizebox{0.46\textwidth}{!}{
\includegraphics{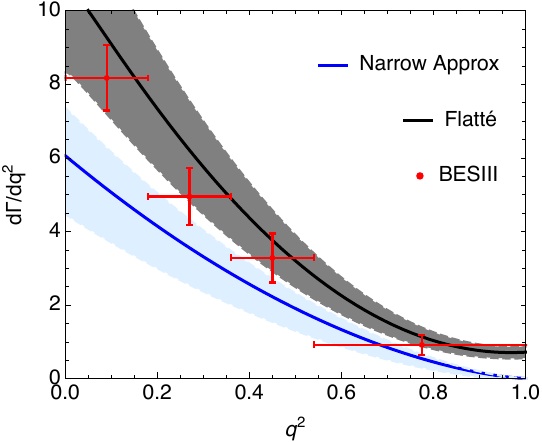}} 
\non
\end{center}
\vspace{-4mm}
\caption{Left: The LCSRs prediction (gray) and the SSE parameterization (lightblue) of $D_s \to f_0$ form factor $f_+(q^2)$. 
Right: The differential width (in unit of ${\rm ns^{-1}/GeV^{2}/c^4}$) of $D_s^+ \to f_0 e^+ \nu_e$ 
obtained from the narrow width approximation (blue) and the Flatt${\rm \acute{e}}$ resonant formula (black). }
\label{fig:Ds2f0-besiii}       
\end{figure}

In the left panel of figure \ref{fig:Ds2f0-besiii}, we depict the $D_s \to f_0$ form factor $f_+(q^2)$ 
obtained by the LCSRs calculation in the large recoiled regions (gray band) 
and the simple $z$-series expansion parameterization (SSE) \cite{Bourrely:2008za} in the small recoiled region (lightblue band). 
For the sake of comparison, we also plot the extracted form factor from the differential width $d\Gamma/dq^2$ 
measured at BESIII with the Flatt\'e model. 
Our prediction is consistent with the data extraction, while shows a litter bit larger. 
In the right panel, we depict the differential width of $D_s^+ \to \left[ \pi\pi \right]_{\rm S} e^+ \nu_e$ on the momentum transfers, 
where the result obtained from the narrow width approximation in Eq. (\ref{eq:decaywidth-massless}) 
and the Flatt\'e resonant model in Eq. (\ref{eq:decaywidth-resonant}) are plotted in blue and black curves, respectively. 
The curve of decay width obtained from the narrow width approximation is a litter bit lower than the data, 
the result with considering the width effect by resonant model is in consistent with the data with showing a litter bit larger. 
This difference indicates the sensitive of differential decay width on the resonant model. 
Based on the discussions, we would like to comment that the extraction of $D_s \to f_0$ form factor by the BESIII measurement of differential decay width, 
as well as our result of decay width from the LCSRs form factor, is model dependent. 
A model independent analysis is highly anticipated to do the high accuracy phenomena study. 

\section{$D_s \to \left[ \pi\pi \right]_{\rm S}$ from factors}\label{sec:Ds2pipi-ff}

A model independent study is to consider directly the stable $\pi\pi$ state on-shell rather than the $f_0$. 
In this case the $D_s(p) \to \left[ \pi(k_1)\pi(k_2) \right]_{\rm S} l(p_2) \nu(p_3)$ decay amplitude can be written as
\beq
{\cal M}(D_s^+ \to \left[ \pi\pi \right]_{\rm S} e^+ \nu) =
 - \frac{i G_F}{\sqrt{2}} V_{cs}^\ast {\bar u}_e(p_2) \gamma^\mu (1 - \gamma_5) v_{\nu_4}(p_3) 
\langle \left[ \pi(k_1)\pi(k_2) \right]_{\rm  S} \vert {\bar s} \gamma_\mu (1 - \gamma_5) c \vert D_s^+(p) \rangle, 
\eeq 
in which the hadron transition matrix element is decomposed in terms of the orthogonal form factors 
\beq
\langle \left[ \pi(k_1)\pi(k_2) \right]_{\rm S} \vert {\bar s} \gamma_\mu (1 - \gamma_5) c \vert D_s^+(p) \rangle 
= - i F_t(q^2, s, \zeta)  k_\mu^t - i F_0(q^2, s, \zeta) k_\mu^0 - i F_\parallel(q^2, s, \zeta) k_\mu^\parallel
\label{eq:Ds2pipi_ff}
\eeq
accompanying with the kinematical variables
\beq
k_\mu^t = \frac{q_\mu}{\sqrt{q^2}}  \,, \quad 
k_\mu^0 = \frac{2\sqrt{q^2}}{\sqrt{\lambda_{D_s}}} \left( k_\mu - \frac{k \cdot q}{q^2} q_\mu \right) \,, \quad 
k_\mu^\parallel = \frac{1}{\sqrt{k^2}} \left( {\bar k}_\mu - \frac{4 (q \cdot k)(q \cdot {\bar k})}{\lambda_{D_s} } k_\mu 
+ \frac{4 k^2 (q \cdot {\bar k})}{\lambda_{D_s}} q_\mu \right). 
\eeq
Besides the momentum transfers $q^2$ in the weak decay and the invariant mass $k^2 \equiv (k_1 + k_2)^2$ of dipion system, 
$D_s \to \left[\pi\pi\right]_{\rm S}$ form factors also carries the information about the angular momentum between two collinear pions 
by the variable $2 q \cdot {\bar k} \equiv 2 q \cdot (k_1 - k_2) = \sqrt{\lambda_{D_s}} (2 \zeta -1) = \sqrt{\lambda_{D_s}} \beta_{\pi\pi}(k^2) \cos \theta_\pi$. 
Here $\theta_\pi$ is the angle between the 3-momenta of $\pi(k_2)$ meson and the $D_s(p)$ meson in the dipion rest frame, 
the Kall\"en function is $\lambda_{D_s} = \lambda(m_{D_s}^2, k^2, q^2)$.

Multiplying Eq. (\ref{eq:Ds2pipi_ff}) by the polarization vector of weak (lepton-neutrino pair) current, we can define the helicity form factors
\beq
H_{i}^{D_s \to \left[\pi\pi \right]_{\rm S}}(q^2, k^2, \zeta) = 
{\bar \epsilon}^\mu(i) \langle \left[ \pi(k_1)\pi(k_2) \right]_{\rm S} \vert {\bar s} \gamma_{\mu} (1 - \gamma_5) c \vert D_s^+(p) \rangle \,.
\eeq
The subscript $i = 0, t$ denotes the polarization direction, 
hereafter we would not show explicitly the superscript for the sake of simplicity. 
We immediately obtain the simple relation between the helicity form factors and the orthogonal form factors, 
\beq
H_{0}(q^2, k^2, \zeta) = - i F_0(q^2, k^2, \zeta), \qquad  H_{t}(q^2, k^2, \zeta) = -i F_t(q^2, k^2, \zeta).
\eeq
We then obtain the three-order differential width of semileptonic decay $D_s^+(p) \to \left[ \pi(k_1)\pi(k_2) \right]_{\rm S} e^+(p_2) \nu_e(p_3)$
\beq
\frac{d^3\Gamma(D_s^+ \to \left[ \pi\pi \right]_{\rm S} l^+ \nu)}{dk^2 dq^2 d(\cos \theta_\pi)} 
&=& \frac{G_F^2 \vert V_{cs}\vert^2}{192 \pi^3 m_{D_s}^3} \frac{\beta_{\pi\pi}(k^2) \sqrt{\lambda_{D_s}} q^2 }{16 \pi} \left[ -i F_0(q^2, k^2, \zeta) \right]^2. 
\label{eq:Ds2pipilv-3rd}
\eeq
The key issue becomes to calculate the $D_s \to \left[ \pi\pi \right]_{\rm S}$ form factors $F_0(q^2, k^2, \zeta)$. 

\subsection{The chiral even generalized $2\pi$ distribution amplitudes}\label{ssec:LCDAs-pipi-t2}

In order to calculate the $D_s \to \left[ \pi\pi \right]_{\rm S}$ form factors, we should firstly know the LCDAs ($2\pi$DAs) of $\pi\pi$ system. 
We introduce a parameter angle to describe the mixing between ${\bar n}n = ({\bar u}u + {\bar d}d)/\sqrt{2}$ 
and ${\bar s}s$ in the isoscalar scalar $\pi\pi$ and $KK$ states
\beq
\left[ \pi\pi \right]_{\rm S} = \vert {\bar n}n \rangle \cos \theta + \vert {\bar s}s \rangle \sin \theta, \qquad
\left[ KK \right]_{\rm S} = - \vert {\bar n}n \rangle \sin \theta + \vert {\bar s}s \rangle \cos \theta.
\eeq
The chiral even two quark isoscalar $2\pi$DAs \cite{Diehl:1998dk,Polyakov:1998ze} involved in our calculation is defined by 
\beq
&&\langle \left[ \pi^a(k_1) \pi^b(k_2)\right]_{\rm S} \vert {\bar s}(xn) \gamma_\mu s(0) \vert 0 \rangle = 2 \delta^{ab} k_\mu  \sin \theta \int du e^{iu x(k \cdot n)} 
\Phi_{\parallel, \left[ \pi\pi\right]_{\rm S}}^{I=0} (u, \zeta, k^2) \,.
\label{eq:2piDAs-def}
\eeq
It is easy to check the $C$-parity symmetry properties 
\beq
\Phi_{\parallel, \left[ \pi\pi\right]_{\rm S}}^{I=0} (u, \zeta, k^2) 
= - \Phi_{\parallel, \left[ \pi\pi\right]_{\rm S}}^{I=0} (1- u, \zeta, k^2) 
= \Phi_{\parallel, \left[ \pi\pi\right]_{\rm S}}^{I=0} (u, 1-\zeta, k^2) \,.
\eeq
To describe the $\pi\pi$ system, three independent kinematical variables are introduced, 
they are the momentum fraction $u$ carried by the antiquark with respect to the total momentum $k$, 
the longitudinal momentum fraction $\zeta = k^{+}_1/k^{+}$ carried by one pion in the system, and the invariant mass square $k^2 = s$. 

The generalized isoscalar scalar $2\pi$DAs is normalized by the quark part of energy momentum tensor form factor $F_\pi^{ \rm EMT}$, 
\beq
\int du (2u - 1) \Phi_{\parallel, \left[ \pi\pi\right]_{\rm S}}^{I=0}(u, \zeta, k^{2}) = - 2 M_2^{(\pi)} \zeta (1-\zeta) F_\pi^{ \rm EMT}(k^{2}),
\label{eq:normalizations}
\eeq
in which $M_2^{(\pi)}$ indicates the second moment of quark distributions in the pion $M_2^{(\pi)} = 2 \int_0^1 du u \left[ q_\pi(u) + {\bar q}_\pi(x) \right]$, 
$F_\pi^{ \rm EMT}(0) = 1$. 

With the double expansion of eigenfunction of the evolution function and the partial wave, 
the generalized $2\pi$DAs is written by the detached Gegenbauer and Legendre polynomials 
\beq
\Phi_{\parallel, \left[ \pi\pi\right]_{\rm S}}^{I}(u, \zeta, k^{2}, \mu) = 6 u (1-u) \sum_{n=1, {\rm odd}}^{\infty} \sum_{l=0, {\rm even}}^{n+1} 
B^{I=0}_{\parallel, nl}(k^{2}, \mu) C_n^{3/2}(2u-1)C_l^{1/2}(2 \zeta -1) \,.
\label{eq:2KDAs-exp}
\eeq
The $k^2$-dependent expansion coefficient $B_{nl}$ have the similar scale evolution as the gegenbauer coefficients in the LCDAs of $f_0$ meson \cite{Cheng:2019hpq}. 
\beq
B^{I=0}_{\parallel, nl}(\mu, k^{2}) = B^{I=0}_{\parallel, nl}(0) \left[ \frac{\alpha_s(\mu)}{\alpha_s(\mu_0)}\right]^{\frac{\gamma_n^{(0)} - \gamma_0^{(0)}}{(2\beta_0)}} 
{\rm Exp} \left[ \sum_{m=1}^{N-1} \frac{k^{2m}}{m!} \frac{d^m}{dk^{2m}} \ln B^{I=0}_{\parallel, nl}(0) 
+ \frac{k^{2N}}{\pi} \int_{4m_K^2}^\infty ds \frac{\delta_l^{I=0}(s)}{s^N (s-k^{2}-i0)} \right].
\label{eq:Bnl}
\eeq 
the pion-pion phase shift $\delta_l^{I=0}(s)$ carries the resonant information in the dipion invariant mass spectral. 
At one-loop level, the anomalous dimension reads as 
\beq
\gamma_n^{\parallel,(0)} = 8 C_F \left( \sum_{k=1}^{n+1} \frac{1}{k} - \frac{3}{4} - \frac{1}{2(n+1)(n+2)} \right), 
\label{eq:ad-1loop}
\eeq
the $\beta$ function coefficient is $\beta_0 = 11 - 2 N_f/3$. 
Based on the Watson theorem of scattering amplitudes, 
the exponential function in Eq. (\ref{eq:Bnl}) is the solution of the $N$-subtracted dispersion relation for the coefficient $B_{nl}$, 
whose evolution could touch up to $\sim 2.5 \, {\rm GeV}$. 

\begin{figure}[tb]
\begin{center}
\resizebox{0.46\textwidth}{!}{
\includegraphics{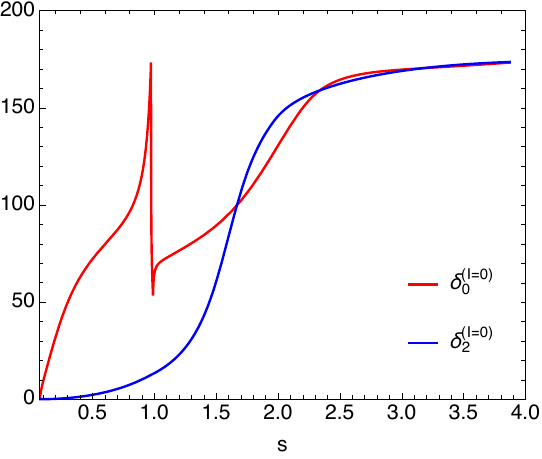}} 
\hspace{8mm}
\resizebox{0.45\textwidth}{!}{
\includegraphics{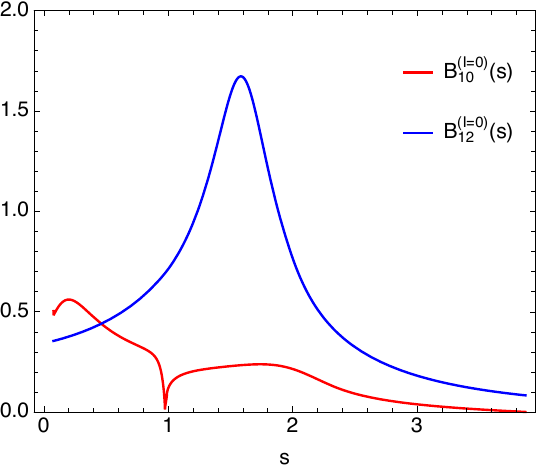}} 
\non
\end{center}
\vspace{-6mm}
\caption{The phase shifts $\delta^{(I=0)}_l(s)$ (left) and the expansion coefficients $B_{1l}^{(I=0)}(s)$ (right)
of isoscalar $\pi\pi$ system in the $S$ and $D$-waves.}
\label{fig:phase-B1012}       
\end{figure}

Concerning the coefficients at the zero point of invariant mass, the soft pion theorem and the crossing symmetry relate it with the 
gegenbauer moments $a_n$ and the moments of quark distribution $M_N$ in pion.
\beq
\sum_{l=0}^{n+1} B^{I=0}_{\parallel, nl} = 0 \,, \quad
B^{I=0}_{\parallel, N-1N}(0) = \frac{1}{3} \frac{2N+1}{N+1} M^{(\pi)}_{N = {\rm even}} \,.
\label{eq:coef-relations}
\eeq
In the vicinity of the resonance, isoscalar scalar $2\pi$DAs deduces to the distribution amplitudes of $f_0$ 
\beq
{\bar f}_{f_0} a_n^{f_0} = B_{\parallel, n0}^{I=0}(0) {\rm Exp} \left[ \sum_{m=1}^{n-1} c_m^{(n0)} m_{f_0}^{2m} \right], \quad 
{\bar f}_{f_0} a_1^{f_0} = \frac{\Gamma_{f_0} {\rm Im} B_{\parallel, 12}^{I=0}(m_{f_0}^2)}{g_{f_0 \pi\pi}}
\label{eq:coef-relations-1}
\eeq
with the subtraction coefficient
\beq
c_m^{(nl)} = \frac{1}{m!} \frac{d^m}{dk^{2m}} \left[ \ln B^{I=0}_{\parallel, nl}(k^2) - \ln B^{I=0}_{\parallel, l+1l}(k^2) \right]_{k^2 \rightarrow 0}.
\label{eq:coef-subtraction}
\eeq
Based on the QCDSR calculation of LCDAS of $f_0$ in the section \ref{sec:LCDAs}, 
we obtain the first expansion and subtraction coefficients of isoscalar scalar dipion system.  
\beq
&&{\bar f}_{f_0} a_1^{f_0} = B^{I=0}_{\parallel, 10}(0) = \frac{\Gamma_{f_0} {\rm Im} B_{\parallel, 10}^{I=0}(m_{f_0}^2)}{g_{f_0 \pi\pi}} = -0.300, \quad 
B^{I=0}_{\parallel, 12} = -B^{I=0}_{\parallel, 10} = 0.300,\non
&&\frac{d \ln B^{I=0}_{\parallel, 10}(k^2)}{d k^2}\vert_{k^2 \to 0} = \frac{d \ln B^{I=0}_{\parallel, 12}(k^2)}{d k^2}\vert_{k^2 \to 0} = \frac{N_c}{48 \pi^2 f_\pi^2} = 
0.375.
\eeq
In figure \ref{fig:phase-B1012}, we depict the phase shift $\delta_{l}^{I=0}(s)$ from the amplitude analysis with a combination of dispersion relations and unitarity \cite{Dai:2014zta,Dai:2017uao}, and the obtained expansion coefficient $B_{1l}^{I=0}(s)$. 
We can clearly see a sharp dip around the $f_0$ region in the $S$-wave contribution to phase shift and hence the first-order expansion coefficient, 
additionally, a quick rising around $f_0(1370)$ region in the $D$-wave contributions.

\subsection{$D_s \to \left[ \pi\pi \right]_{\rm S}$ from factor $F_0^{(l)}(q^2, s)$ at leading twist}\label{ssec:Ds2pipi-ff-t2}

To evaluate the $D_s \to \left[ \pi\pi \right]_{\rm S}$ form factors, we consider the nonlocal correlation function
\beq
&&\Pi_\mu^{ab}(q, k_1, k_2) = i \int d^4x e^{iq \cdot x} \langle \pi^a(k_1) \pi^b(k_2) \vert T \{j_{1, \mu}^{\rm S1}(x), j_{2}^{\rm S1}(0) \} \vert 0 \rangle,
\label{eq:correlation1}
\eeq
and  take the $D_s$ interpolating current and weak decay currents as the same as in the scenario I (${\rm S1}$) in the calculation of $D_s \to f_0$ form factors. 
Since our knowledge of $2\pi$DAs is still at leading twist level so far, we can only discuss the width effect of $D_s \to \left[\pi\pi \right]_{\rm S}$ at leading twist.  
We furthermore consider an auxiliary correlation function 
\beq
\Pi^{ab}(q, k_1, k_2) = i \int d^4x e^{iq \cdot x} \langle \pi^a(k_1) \pi^b(k_2) \vert T \{j_5(x), j_{2}^{\rm S1}(0) \} \vert 0 \rangle \,.
\label{eq:correlation2}
\eeq 
to evaluate the timelike helicity form factor $F_t$ with the weak current $j_5 = - i m_c {\bar s} \gamma_5 c$. 
The auxiliary correlation function can be obtained by multiplying Eq. (\ref{eq:correlation1}) with $q_\mu$. 

For the sake of simplicity, we take the neutral dipion system with electric charge $a=b=0$ for example to show the LCSRs evaluation. 
The correction functions in Eqs. (\ref{eq:correlation1},\ref{eq:correlation2}) can be written in the hadron representation as 
\beq
\Pi_\mu^{\rm had}(q, k_1, k_2) &=& \frac{im_{D_s}^2 f_{D_s}}{\left[ m_{D_s}^2 - (k+q)^2 \right] (m_c + m_s)} 
\left[ F_t(q^2,k^2,\zeta) k_\mu^t + F_0(q^2,k^2,\zeta) k_\mu^0 + F_\parallel(q^2,k^2,\zeta) k_\mu^\parallel \right] \non
&+& \frac{1}{\pi} \int_{s_0}^\infty ds \frac{\rho_t^h(q^2,s) k_\mu^t + \rho_0^h(q^2,s) k_\mu^0 + \rho_\parallel^h(q^2,s) k_\mu^\parallel}{s - (k+q)^2} \,, 
\label{eq:correlation1-hadron}\\ 
\Pi^{\rm had}(q, k_1, k_2) &=& \frac{m_{D_s}^2 f_{D_s}}{\left[ m_{D_s}^2 - (k+q)^2 \right] (m_c + m_s)} \left[\sqrt{q^2} F_t(q^2,k^2,\zeta) \right] 
+ \frac{1}{\pi} \int_{s_0}^\infty ds \frac{\rho^h(q^2,s)}{s - (k+q)^2} \,.
\label{eq:correlation2-hadron}
\eeq
Meanwhile, the OPE calculation of these correlation functions result in 
\beq
\Pi_\mu^{\rm OPE}(q, k_1, k_2) &=& 2 i \sin\theta m_c k_\mu \int_0^1 \frac{du}{u} 
\frac{\Phi_{\parallel, \left[ \pi\pi\right]_{\rm S}}^{I=0, s{\bar s}}(u, \zeta, k^2)}{s^\prime_2(u) - (k+q)^2} \,, 
\label{eq:correlation1-OPE}\\ 
\Pi^{\rm OPE}(q, k_1, k_2) &=& \frac{m_c \sin\theta }{2} \int_0^1 \frac{du}{u} 
\frac{\Phi_{\parallel, \left[ \pi\pi\right]_{\rm S}}^{I=0, s{\bar s}}(u, \zeta, k^2) \left[ m_{D_s}^2 - q^2 - (1-2u) k^2 \right]}{s^\prime_2(u) - (k+q)^2} \,. 
\label{eq:correlation2-OPE}
\eeq
After applying the quark-hadron duality and Borel transformation, we obtain the relations between $D_s \to \left[ \pi\pi \right]_{\rm S}$ form factors
\beq
&&\frac{2 \sqrt{q^2}}{\lambda_{D_s}} F_0(q^2, k^2, \zeta) - \frac{4(q \cdot k)(q \cdot {\bar k})}{\lambda_{D_s} \sqrt{k^2}} F_\parallel(q^2, k^2, \zeta) 
= \frac{2 m_c (m_c + m_s) \sin\theta}{m_{D_s}^2 f_{D_s}} \int_{u_0}^1 \frac{du}{u} 
\, \Phi_{\parallel, \left[ \pi\pi\right]_{\rm S}}^{I=0, s{\bar s}}(u, \zeta, k^2) \, e^{- \frac{s^\prime_2(u) - m_{D_s}^2}{M^2}}\,,
\label{eq:correlation-SR1}\\ 
&&\frac{F_t(q^2, k^2, \zeta)}{\sqrt{q^2}} - \frac{2 (q \cdot k) F_0(q^2, k^2, \zeta)}{\sqrt{\lambda_{D_s} q^2} } 
+ \frac{4 \sqrt{ k^2} (q \cdot {\bar k}) F_\parallel(q^2, k^2, \zeta)}{\lambda_{D_s}} = 0 \,.
\label{eq:correlation-SR2}
\eeq
Here $s^\prime_2(u) = {\bar u} k^2 + (m_c^2 - {\bar u}q^2)/u$. 
The sum rules are ultimately obtained as 
\beq
&&\cos \theta_\pi F_\parallel(q^2, k^2, \zeta) = \frac{m_c(m_c+m_s) \sin\theta }{m_{D_s}^2 f_{D_s} \sqrt{\lambda_{D_s}} \beta_{\pi\pi}(k^2) } 
\int_{u_0}^1 \frac{du}{u} \, \left[ 4 u \left( k^2 \right)^{3/2} \right] 
\Phi_{\parallel, \left[ \pi\pi\right]_{\rm S}}^{I=0, s{\bar s}}(u, \zeta, k^2) \, e^{- \frac{s^\prime_2(u) - m_{D_s}^2}{M^2}},
\label{eq:Ds2pipi-SR1}\\
&&F_0(q^2, k^2, \zeta) = \frac{m_c(m_c+m_s) \sin\theta}{m_{D_s}^2 f_{D_s} \sqrt{\lambda_{D_s}} \sqrt{q^2} } 
\int_{u_0}^1 \frac{du}{u} \, \left[ \lambda_{D_s} + 2u k^2 \left( m_{D_s}^2 + q^2 - k^2 \right) \right] 
\Phi_{\parallel, \left[ \pi\pi\right]_{\rm S}}^{I=0, s{\bar s}}(u, \zeta, k^2) \, e^{- \frac{s^\prime_2(u) - m_{D_s}^2}{M^2}},
\label{eq:Ds2pipi-SR2}\\
&&\sqrt{q^2} F_t(q^2, k^2, \zeta) = \frac{m_c (m_c + m_s) \sin\theta}{m_{D_s}^2 f_{D_s}} \int_{u_0}^1 \frac{du}{u} 
\,\left[ m_{D_s}^2 - q^2 - (1-2u)k^2 \right] \Phi_{\parallel, \left[ \pi\pi\right]_{\rm S}}^{I=0, s{\bar s}}(u, \zeta, k^2) \, e^{- \frac{s^\prime_2(u) - m_{D_s}^2}{M^2}}.
\label{eq:Ds2pipi-SR3}
\eeq 
We can check that $F_0(q^2, k^2, \zeta) = F_t(q^2, k^2, \zeta)$ at the full recoiled point $q^2 = 0$.  
From the view of partial-wave analysis \cite{Faller:2013dwa}, $D_s \to \pi\pi$ form factors are expanded by 
\beq
&&F_{0,t}(q^2, k^2, \zeta) = \sum_{\ell=0}^\infty \sqrt{2\ell+1} \, F_{0,t}^{(\ell)}(q^2, k^2) P_\ell^{(0)}(\cos \theta_\pi) \,, \non
&&F_{\parallel,\perp}(q^2, k^2, \zeta) = \sum \sum_{\ell = 0} \sqrt{2\ell+1} \, 
F_{\parallel,\perp}^{(\ell)}(q^2,k^2) \frac{P_\ell^{(1)}(\cos \theta_\pi)}{\sin \theta_\pi} \,. 
\label{eq:Ds2pipi-ff-wave}
\eeq
The associated Legendre polynomials have the orthogonality relations
\beq
\int_{-1}^1 P_\ell^n(x) P_k^n(x) dx = \frac{2}{2 \ell +1} \frac{(\ell +n)!}{(\ell - n)!} \delta_{k \ell}, \quad
\int_{-1}^1 \frac{P_\ell^m(x) P_\ell^n(x)}{1-x^2} dx = \frac{(\ell + m)!}{m (\ell - m)!} \delta_{mn}\, \quad {\rm with} \, m,n \neq 0 . 
\label{eq:Legendre-orthor}
\eeq

Multiplying $P_\ell^{(0)}(\cos\theta_\pi)$ to both sides of Eqs. (\ref{eq:Ds2pipi-SR1},\ref{eq:Ds2pipi-SR2}) and integrating over $\cos\theta_\pi$, 
we obtain the sum rules of $D_s \to \left[ \pi\pi \right]_{\rm S}$ form factors at $\ell^\prime$-wave ($\ell^\prime = {\rm even}$ and $\ell^\prime \leqslant n+1$)
\beq
&&\sum_{\ell=1}^\infty I_{\ell^\prime \ell}^{I=0} \, F_\parallel^{(\ell)}(q^2, k^2) 
= \frac{m_c(m_c+m_s) \sin\theta}{m_{D_s}^2f_{D_s}\sqrt{\lambda_{Ds}}} 
\sum_{n=1,{\rm odd}}^\infty \frac{1}{2 \ell^\prime + 1} \, J_n^\parallel(q^2,k^2,M^2,s_0) \, B_{n \ell^\prime, \parallel}^{I=0}(k^2),
\label{eq:eq:Ds2pipi-SR1-l}\\
&&F_0^{(\ell^\prime)}(q^2,k^2) = \frac{m_c(m_c+m_s) \sin\theta}{m_{D_s}^2f_{D_s}\sqrt{\lambda_{Ds}}\sqrt{q^2}} 
\sum_{n=1,{\rm odd}}^\infty \frac{\beta_{\pi\pi}(k^2) }{\sqrt{2 \ell^\prime + 1}} \, J_n^0(q^2,k^2,M^2,s_0) \, B_{n \ell^\prime, \parallel}^{I=0}(k^2),
\label{eq:eq:Ds2pipi-SR1-2}\\
&&F_t^{(\ell^\prime)}(q^2,k^2) = \frac{m_c(m_c+m_s) \sin\theta}{m_{D_s}^2f_{D_s}\sqrt{q^2}} 
\sum_{n=1,{\rm odd}}^\infty \frac{\beta_{\pi\pi}(k^2)}{\sqrt{2 \ell^\prime + 1}} \, J_n^t(q^2,k^2,M^2,s_0) \, B_{n \ell^\prime, \parallel}^{I=0}(k^2),
\label{eq:eq:Ds2pipi-SR1-3}
\eeq 
with the conformal expansion functions $J_n$
\beq
&&J_n^\parallel(q^2,k^2,M^2,s_0) = 6 \int_{u_0}^1 du \, {\bar u} \, C_n^{3/2}(2u-1) \left[ 4 \left( k^2 \right)^{3/2} \right] 
e^{- \frac{s^\prime_2(u) - m_{D_s}^2}{M^2}}, \\
&&J_n^0(q^2,k^2,M^2,s_0) = 6 \int_{u_0}^1 du \, {\bar u} \, C_n^{3/2}(2u-1) 
\left[ \lambda_{D_s} + 2u k^2 \left( m_{D_s}^2 + q^2 - k^2 \right) \right] e^{- \frac{s^\prime_2(u) - m_{D_s}^2}{M^2}}, \\
&&J_n^t(q^2,k^2,M^2,s_0) = 6 \int_{u_0}^1 du \, {\bar u} \, C_n^{3/2}(2u-1) \left[m_{D_s}^2 - q^2 - (1-2u) k^2 \right] 
e^{- \frac{s^\prime_2(u) - m_{D_s}^2}{M^2}}.
\eeq
and the additional partial-wave expansion function $I_{\ell^\prime \ell}$
\beq
&&I_{\ell^\prime \ell}^{I=0} = \sqrt{2l+1} \, \int_{-1}^1 d (\cos\theta_\pi)  \frac{\cos\theta_\pi}{\sin\theta_\pi} \, 
P_{\ell^{\prime\prime}}^{(0)}(\cos\theta_\pi) P_{\ell}^{(1)}(\cos\theta_\pi) \,.
\eeq
We mark that $I_{\ell \ell^\prime}^{I=0}$ is zero when $\ell$  goes over odd number, 
$I_{02}^{I=0} = -2 \sqrt{5}, I_{22}^{I=0} = -4/\sqrt{5}$ and $I_{\ell^\prime 2}^{I=0} = 0$ when $\ell^\prime  > 2$. 

\begin{figure}[th]
\begin{center}
\resizebox{0.45\textwidth}{!}{
\includegraphics{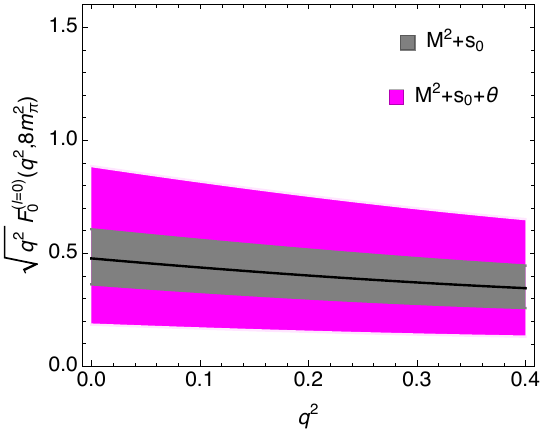}} 
\hspace{8mm}
\resizebox{0.45\textwidth}{!}{
\includegraphics{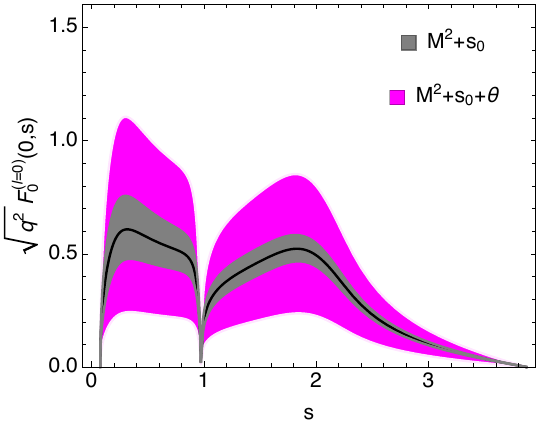}} 
\non
\end{center}
\vspace{-6mm}
\caption{Evolution of the $S$-wave orthogonal $D_s \to \left[\pi\pi \right]_{\rm S}$ form factor on the momentum transfers $\sqrt{q^2}F_{0}^{(l=0)}(q^2, s = 8 m_\pi^2)$ (left) 
and and on the invariant mass $\sqrt{q^2}F_{0}^{(l=0)}(q^2 = 0, s)$ (right).}
\label{fig:F0l0}       
\end{figure}

\begin{figure}[tb]
\begin{center}
\resizebox{0.45\textwidth}{!}{
\includegraphics{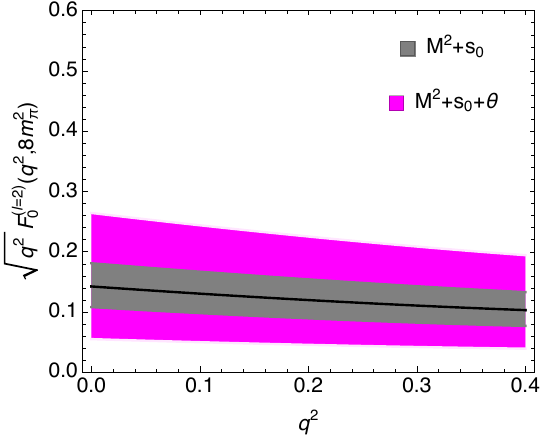}} 
\hspace{8mm}
\resizebox{0.45\textwidth}{!}{
\includegraphics{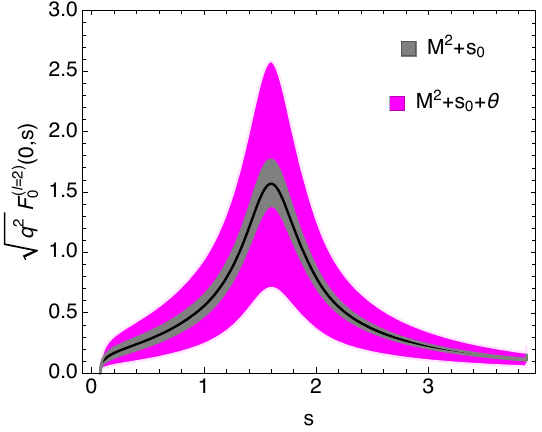}} 
\non
\end{center}
\vspace{-6mm}
\caption{The same as figure. \ref{fig:F0l0}, but for the $D$-wave $\pi\pi$ system.}
\label{fig:F0l2}       
\end{figure}

In figure \ref{fig:F0l0} we depict the evolutions of $S$-wave $D_s \to \left[ \pi\pi \right]_{\rm S}$ form factor 
on the momentum transfers (left panel) and invariant mass (right panel). 
and in figure \ref{fig:F0l2} for the form factor with $D$-wave $\pi\pi$ system. 
We take the mixing angle between the isoscalar scalar $\pi\pi$ and $KK$ systems at $\theta = 20\degree \pm 10\degree$ 
which is similar as the angle in the $\sigma$-$f_0$ mixing \cite{Gokalp:2004ny}. 
The uncertainty arose from the mixing angle is added up to the LCSRs parameters uncertainty and shown in the magnate bands. 
We find that the $D$-wave form factor $\sqrt{q^2}F_{0}^{(l=2)}(q^2)$ is much smaller than the $S$-wave $\sqrt{q^2}F_{0}^{(l=0)}(q^2)$ 
when the invariant mass is small, while in the resonant regions $D$-wave contribution is comparable or even larger than the $S$-wave. 


\subsection{$D_s \to \left[ \pi\pi \right]_{\rm S} e \nu_e$ decay at leading twist}\label{ssec:stable}

\begin{figure}[th]
\vspace{4mm}
\begin{center}
\resizebox{0.45\textwidth}{!}{
\includegraphics{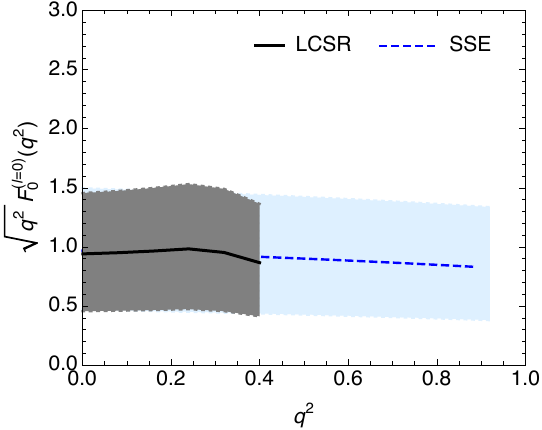}} 
\hspace{8mm}
\resizebox{0.45\textwidth}{!}{
\includegraphics{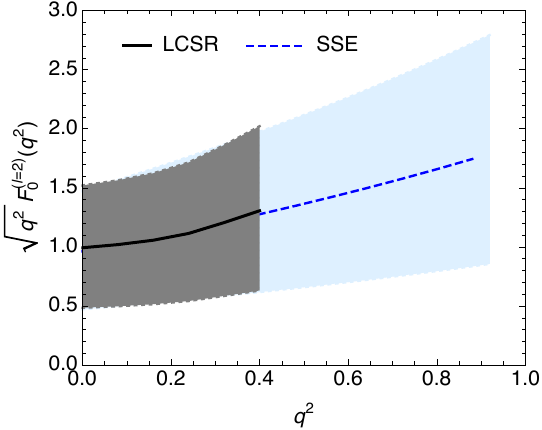}} 
\non
\end{center}
\vspace{-6mm}
\caption{The integrated form factor $\sqrt{q^2} F^{(l)}_{0}(q^2)$ for the $S$-wave (left) and $D$-wave (right) $\pi\pi$ system.}
\label{fig:F0}       
\end{figure}

In figure \ref{fig:F0}, we depict the evolution of form factor on the momentum transfers $q^2$ where the invariant mass dependence is integrated out. 
\beq
\sqrt{q^2} F^{(l)}_{0}(q^2) = \int_{4m_\pi^2}^{s_{\rm max}(q^2)} ds \sqrt{q^2} F^{(l)}_{0}(q^2, s),  
\eeq
here $s_{\rm max}(q^2)$ is the solution of $\lambda(m_{D_s}^2, q^2, s) = 0$. 
Again, the kinematical region with small recoling is extrapolated by the SSE parameterization. 

\begin{figure}[t]
\begin{center}
\resizebox{0.46\textwidth}{!}{
\includegraphics{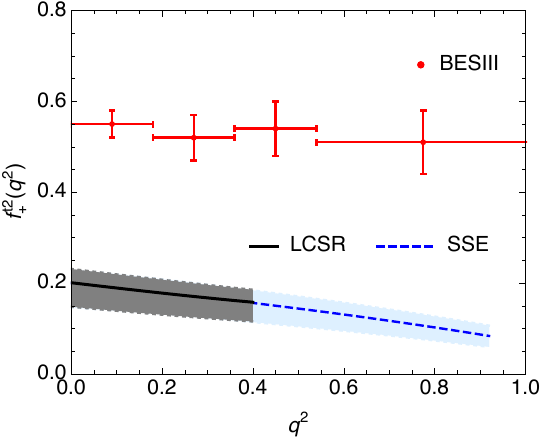}} 
\hspace{6mm}
\resizebox{0.45\textwidth}{!}{
\includegraphics{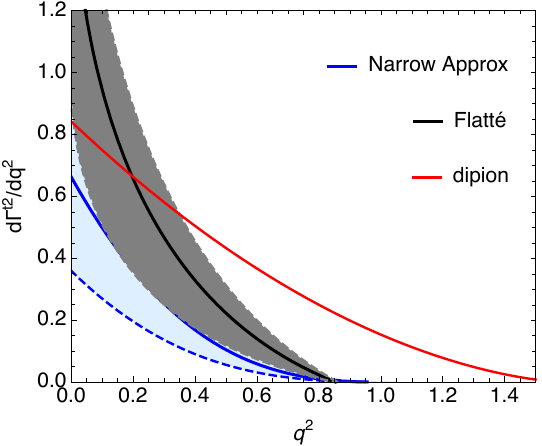}} 
\non
\end{center}
\vspace{-6mm}
\caption{The same as figure. \ref{fig:Ds2f0-besiii}, but for the leading twist contribution. }
\label{fig:Ds2f0-t2}       
\end{figure}

Considering the partial-wave expansion of $D_s \to \pi\pi$ form factors in Eq. (\ref{eq:Ds2pipi-ff-wave}) and the orthogonal conditions in Eq. (\ref{eq:Legendre-orthor}), 
the three-order differential decay width in Eq. (\ref{eq:Ds2pipilv-3rd}) deduces to the second-order one after integrating over the angle $\theta_\pi$, 
\beq
\frac{d^2\Gamma(D_s^+ \to \left[ \pi\pi \right]_{\rm S} l^+ \nu)}{dk^2 dq^2} 
&=& \frac{G_F^2 \vert V_{cs}\vert^2}{192 \pi^3 m_{D_s}^3} \frac{\beta_{\pi\pi}(k^2) \sqrt{\lambda_{D_s}}}{16 \pi} 
\sum_{\ell=0}^\infty 2 \vert \sqrt{q^2} F_0^{(\ell)}(q^2, k^2) \vert^2.
\label{eq:Ds2pipilv-2nd}
\eeq
After doing the integration on the invariant mass, we plot the momentum transfer dependence of the 
$D_s^+ \to \left[ \pi\pi \right]_{\rm S} e^+ \nu_e$ decay width (in unit of ${\rm ns^{-1}/GeV^{2}/c^4}$) in the right panel of figure. \ref{fig:Ds2f0-t2}. 
We mark that the result is obtained at leading twist level of the the dipion LCDAs, 
so we compare it with the result obtained from the narrow width approximation and Flatt\'e resonant model with the leading twist $D_s \to f_0$ form factor, 
which is shown in the left panel. 
The three curves of central value indicates from one side the sensitivity of predictions to the resonant model, 
and from another side that the direct calculation from $D_s \to \left[ \pi\pi \right]$ form factor 
shows relatively moderate evolution with larger allowed momentum transfers.  
We mark again the subleading twist contribution is the dominate one in $D_s \to f_0$ form factors,  
so the further study on the twist three dipion LCDAs could provide a model independent solution to the four-body semileptonic decays of $D_s$ meson. 

\section{summary}\label{sec:summary}

In this work we firstly update the QCD sum rules predictions of isoscalar scalar meson $f_0$, 
especially for the decay constant and the gegenbauer expansion coefficients in the LCDAs, 
followed which we calculate the $D_s \to f_0$ transition form factors from the LCSRs with $f_0$ LCDAs 
and obtain the differential width of semileptonic decay $D_s^+ \to f_0 e^+ \nu_e$. 
With the updated decay constant of $f_0$ which is about two times of the one used in the previous LCSRs calculation, 
our result of $D_s \to f_0$ form factors $f_+(q^2)$ with considering the mixing between ${\bar s}s$ and ${\bar u}u+{\bar d}d$ of $f_0$ 
is in consistent with the data extracted from the measurement of differential width in the whole momentum transfer region, 
indicating that the energetic picture of $f_0$ in charm meson decay is still reliable. 
The result of differential width obtained under the narrow width approximation shows a litter bit lower than the data, 
the result obtained under the intermediate resonant model with Flatt\'e formula shows the consistence. 
In order to get rid of the model dependence and give more powerful prediction, 
we suggest to describe the unstable scalar meson by the dipion LCDAs and calculate the $D_s \to [\pi\pi]_{\rm S}$ form factors. 
At leading twist, the directly calculation of differential width of $D_s^+ \to [\pi\pi]_{\rm S}e^+\nu_e$ decay exhibits a moderate evolution 
on the momentum transfers, comparing to the result obtained under the narrow width approximation and the Flatt\'e model.  

Our calculation of $D_s \to [\pi\pi]_{\rm S}$ form factors is carried out at leading twist due to the finite knowledge of dipion LCDAs, 
so an important issue of further development in this project is to construct the twist three dipion LCDAs and take in to account their contributions. 
Of course, the next-to-leading-order QCD radiation corrections of the correlation function is also imperative to improve the prediction accuracy. 
This work reveals a bright prospect to study the four-body leptonic decays of heavy mesons with the dimeson light-cone distribution amplitudes, 
the future experiment with larger integrated luminosity \cite{Belle-II:2018jsg,LHCb:2020hpv} 
would help us to understand the LCDAs of dipion system much better.

\section{acknowledgements}
We are grateful to Ling-Yun Dai and Hai-Bo Li for fruitful discussions, especially to Hai-Yang Cheng for the careful reading of the draft and the helpful comments. 
SC is supported by the National Science Foundation of China (NSFC) under Grant No. 11975112. 
SLZ acknowledge the support from the Natural Science Foundation of Hunan Province, China under Contract No. 2021JJ40036
and the Fundamental Research Funds for the Central Universities under Contract No. 020400/531118010467.


\begin{thebibliography}{99}

\bibitem{El-Bennich:2008rkp}
B.~El-Bennich, O.~Leitner, J.~P.~Dedonder and B.~Loiseau,
Phys. Rev. D \textbf{79}, 076004 (2009)

\bibitem{Offen:2013nma}
N.~Offen, F.~A.~Porkert and A.~Sch\"afer,
Phys. Rev. D \textbf{88}, no.3, 034023 (2013)

\bibitem{Bediaga:2003hr}
I.~Bediaga and M.~Nielsen,
Phys. Rev. D \textbf{68}, 036001 (2003)

\bibitem{Cheng:2005nb}
H.~Y.~Cheng, C.~K.~Chua and K.~C.~Yang,
Phys. Rev. D \textbf{73}, 014017 (2006)

\bibitem{DeFazio:2001uc}
F.~De Fazio and M.~R.~Pennington,
Phys. Lett. B \textbf{521} (2001), 15-21

\bibitem{Ke:2009ed}
H.~W.~Ke, X.~Q.~Li and Z.~T.~Wei,
Phys. Rev. D \textbf{80}, 074030 (2009)

\bibitem{Aliev:2007uu}
T.~M.~Aliev and M.~Savci,
EPL \textbf{90}, no.6, 61001 (2010)

\bibitem{BESIII:2021tfk}
M.~Ablikim \textit{et al.} [BESIII],
Phys. Rev. D \textbf{103}, no.9, 092004 (2021)

\bibitem{Jaffe:1976ig}
R.~L.~Jaffe,
Phys. Rev. D \textbf{15} (1977), 267

\bibitem{Agaev:2017cfz}
S.~S.~Agaev, K.~Azizi and H.~Sundu,
Phys. Lett. B \textbf{781} (2018), 279-282

\bibitem{Weinstein:1982gc}
J.~D.~Weinstein and N.~Isgur,
Phys. Rev. Lett. \textbf{48} (1982), 659

\bibitem{Weinstein:1983gd}
J.~D.~Weinstein and N.~Isgur,
Phys. Rev. D \textbf{27} (1983), 588

\bibitem{Weinstein:1990gu}
J.~D.~Weinstein and N.~Isgur,
Phys. Rev. D \textbf{41} (1990), 2236

\bibitem{CLEO:2009dyb}
J.~Yelton \textit{et al.} [CLEO],
Phys. Rev. D \textbf{80}, 052007 (2009)

\bibitem{CLEO:2009ugx}
K.~M.~Ecklund \textit{et al.} [CLEO],
Phys. Rev. D \textbf{80}, 052009 (2009)

\bibitem{Hietala:2015jqa}
J.~Hietala, D.~Cronin-Hennessy, T.~Pedlar and I.~Shipsey,
Phys. Rev. D \textbf{92}, no.1, 012009 (2015)

\bibitem{BESIII:2021drk}
M.~Ablikim \textit{et al.} [BESIII],
Phys. Rev. D \textbf{105}, no.3, L031101 (2022)

\bibitem{BESIII:2023wgr}
M.~Ablikim \textit{et al.} [BESIII],
[arXiv:2303.12927 [hep-ex]]

\bibitem{Govaerts:1986ua}
J.~Govaerts, L.~J.~Reinders, F.~de Viron and J.~Weyers,
Nucl. Phys. B \textbf{283}, 706-722 (1987)

\bibitem{Yang:1993bp}
K.~C.~Yang, W.~Y.~P.~Hwang, E.~M.~Henley and L.~S.~Kisslinger,
Phys. Rev. D \textbf{47}, 3001-3012 (1993)

\bibitem{Lu:2006fr}
C.~D.~Lu, Y.~M.~Wang and H.~Zou,
Phys. Rev. D \textbf{75}, 056001 (2007)

\bibitem{Ioffe:2002ee}
B.~L.~Ioffe,
Phys. Atom. Nucl. \textbf{66}, 30-43 (2003)

\bibitem{Wu:2022qqx}
Z.~H.~Wu, H.~B.~Fu, T.~Zhong, D.~Huang, D.~D.~Hu and X.~G.~Wu,
[arXiv:2211.05390 [hep-ph]]

\bibitem{Cheng:2019tgh}
S.~Cheng and J.~M.~Shen,
Eur. Phys. J. C \textbf{80}, no.6, 554 (2020)

\bibitem{Han:2013zg}
H.~Y.~Han, X.~G.~Wu, H.~B.~Fu, Q.~L.~Zhang and T.~Zhong,
Eur. Phys. J. A \textbf{49}, 78 (2013)

\bibitem{Wang:2008da}
Y.~M.~Wang, M.~J.~Aslam and C.~D.~Lu,
Phys. Rev. D \textbf{78} (2008), 014006

\bibitem{Braun:2003rp}
V.~M.~Braun, G.~P.~Korchemsky and D.~M\"uller,
Prog. Part. Nucl. Phys. \textbf{51}, 311-398 (2003)

\bibitem{Bijnens:2002mg}
J.~Bijnens and A.~Khodjamirian,
Eur. Phys. J. C \textbf{26}, 67-79 (2002)

\bibitem{Ball:2006wn}
P.~Ball, V.~M.~Braun and A.~Lenz,
JHEP \textbf{05}, 004 (2006)

\bibitem{Colangelo:2010bg}
P.~Colangelo, F.~De Fazio and W.~Wang,
Phys. Rev. D \textbf{81}, 074001 (2010)

\bibitem{Sun:2010nv}
Y.~J.~Sun, Z.~H.~Li and T.~Huang,
Phys. Rev. D \textbf{83}, 025024 (2011)

\bibitem{Gokalp:2004ny}
A.~Gokalp, Y.~Sarac and O.~Yilmaz,
Phys. Lett. B \textbf{609}, 291-297 (2005)

\bibitem{Fleischer:2011au}
R.~Fleischer, R.~Knegjens and G.~Ricciardi,
Eur. Phys. J. C \textbf{71} (2011), 1832

\bibitem{Cheng:2022vbw}
H.~Y.~Cheng, C.~W.~Chiang and Z.~Q.~Zhang,
Phys. Rev. D \textbf{105} (2022) no.3, 033006

\bibitem{LHCb:2013dkk}
R.~Aaij \textit{et al.} [LHCb],
Phys. Rev. D \textbf{87} (2013) no.5, 052001

\bibitem{Khodjamirian:2000ds}
A.~Khodjamirian, R.~Ruckl, S.~Weinzierl, C.~W.~Winhart and O.~I.~Yakovlev,
Phys. Rev. D \textbf{62}, 114002 (2000)

\bibitem{Cheng:2022mvd}
S.~Cheng, Y.~h.~Ju, Q.~Qin and F.~s.~Yu,
Eur. Phys. J. C \textbf{82}, no.11, 1037 (2022)

\bibitem{Duplancic:2008ix}
G.~Duplancic, A.~Khodjamirian, T.~Mannel, B.~Melic and N.~Offen,
JHEP \textbf{04} (2008), 014

\bibitem{Li:2008tk}
R.~H.~Li, C.~D.~Lu, W.~Wang and X.~X.~Wang,
Phys. Rev. D \textbf{79}, 014013 (2009)

\bibitem{Ghahramany:2009zz}
N.~Ghahramany and R.~Khosravi,
Phys. Rev. D \textbf{80}, 016009 (2009)

\bibitem{BES:2004twe}
M.~Ablikim \textit{et al.} [BES],
Phys. Lett. B \textbf{607}, 243-253 (2005)

\bibitem{Bourrely:2008za}
C.~Bourrely, I.~Caprini and L.~Lellouch,
Phys. Rev. D \textbf{79}, 013008 (2009) [erratum: Phys. Rev. D \textbf{82}, 099902 (2010)]

\bibitem{Diehl:1998dk}
M.~Diehl, T.~Gousset, B.~Pire and O.~Teryaev,
Phys. Rev. Lett. \textbf{81}, 1782-1785 (1998)

\bibitem{Polyakov:1998ze}
M.~V.~Polyakov,
Nucl. Phys. B \textbf{555}, 231 (1999)

\bibitem{Cheng:2019hpq}
S.~Cheng,
Phys. Rev. D \textbf{99}, no.5, 053005 (2019)

\bibitem{Dai:2014zta}
L.~Y.~Dai and M.~R.~Pennington,
Phys. Rev. D \textbf{90}, no.3, 036004 (2014)

\bibitem{Dai:2017uao}
L.~Y.~Dai and U.~G.~Mei\ss{}ner,
Phys. Lett. B \textbf{783}, 294-300 (2018)

\bibitem{Faller:2013dwa}
S.~Faller, T.~Feldmann, A.~Khodjamirian, T.~Mannel and D.~van Dyk,
Phys. Rev. D \textbf{89}, no.1, 014015 (2014)

\bibitem{Belle-II:2018jsg}
E.~Kou \textit{et al.} [Belle-II],
PTEP \textbf{2019}, no.12, 123C01 (2019)[erratum: PTEP \textbf{2020}, no.2, 029201 (2020)]

\bibitem{LHCb:2020hpv}
R.~Aaij \textit{et al.} [LHCb],
JHEP \textbf{12}, 144 (2020)
\end{thebibliography}
\end{document}